\documentclass{aastex6}
\slugcomment{Printed \today}
\shorttitle{Non-Thermal Jet from OMC-2 FIR 3}
\shortauthors{Osorio et al.}

\begin{document}


\title{
Star Formation Under the Outflow: The Discovery of a Non-Thermal Jet 
from OMC-2 FIR 3 and its Relationship to the Deeply Embedded FIR 4 
Protostar
}



\author{Mayra Osorio\altaffilmark{1}}
\affil{Instituto de Astrof\'\i sica de Andaluc\'\i a (CSIC) \\ 
Glorieta de la Astronom\'\i a s/n \\
E-18008 Granada, Spain}
\author{Ana K. D\'\i az-Rodr\'\i guez} 
\affil{Instituto de Astrof\'\i sica de Andaluc\'\i a (CSIC) \\ 
Glorieta de la Astronom\'\i a s/n \\
E-18008 Granada, Spain}
\author{Guillem Anglada}
\affil{Instituto de Astrof\'\i sica de Andaluc\'\i a (CSIC) \\ 
Glorieta de la Astronom\'\i a s/n \\
E-18008 Granada, Spain}
\author{S. Thomas Megeath}
\affil{Ritter Astrophysical Research Center,\\
Department of Physics and Astronomy\\
University of Toledo \\
2801 West Bancroft Street\\
Toledo, OH 43606, USA}
\author{Luis F. Rodr\'\i guez}
\affil{Instituto de Radioastronom\'{\i}a y Astrof\'{\i}sica, UNAM \\
Apartado Postal 3-72 (Xangari), 58089 Morelia, Michoac\'an, Mexico} 
\author{John J. Tobin\altaffilmark{2}}
\affil{Homer L. Dodge Department of Physics and Astronomy \\
University of Oklahoma, Norman, OK, USA}
\author{Amelia M. Stutz}
\affil{Department of Astronomy, University of Concepci\' on \\
Concepci\' on, Chile}
\author{Elise Furlan}
\affil{IPAC, Mail Code 314-6, Caltech \\ 
1200 E. California Blvd., Pasadena, CA 91125, USA}
\author{William J. Fischer}
\affil{Space Telescope Science Institute \\ 3700 San Martin Drive, 
Baltimore, MD 21218, USA}
\author{P. Manoj}
\affil{Tata Institute of Fundamental Research \\
Homi Bhabha Road, Mumbai 400 005, India} 
\author{Jos\'e F. G\'omez}
\affil{Instituto de Astrof\'\i sica de Andaluc\'\i a (CSIC) \\
Glorieta de la Astronom\'\i a s/n \\
E-18008 Granada, Spain}
\author{Beatriz Gonz\'alez-Garc\'\i a\altaffilmark{3}}
\affil{European Space Astronomy Center, ESA \\ 
P.O. Box 78, 28691 Villanueva de la Ca\~nada, Madrid, Spain}
\author{Thomas Stanke}
\affil{European Southern Observatory \\ 
Garching bei  M\"unchen, Germany}
\author{Dan M.  Watson}
\affil{Department of Physics and Astronomy, University of Rochester \\ 
Rochester, NY 14627, USA}
\author{Laurent Loinard\altaffilmark{4}}
\affil{Instituto de Radioastronom\'{\i}a y Astrof\'{\i}sica, UNAM \\ 
Apartado Postal 3-72 (Xangari), 58089 Morelia, Michoac\'an, Mexico} 
\author{Roland Vavrek}
\affil{European Space Astronomy Center, ESA \\ 
P.O. Box 78, 28691 Villanueva de la Ca\~nada, Madrid, Spain}

\and

\author{Carlos Carrasco-Gonz\'alez}
\affil{Instituto de Radioastronom\'{\i}a y Astrof\'{\i}sica, UNAM  \\
Apartado Postal 3-72 (Xangari), 58089 Morelia,
Michoac\'an, Mexico} 

\altaffiltext{1}{osorio@iaa.es}

\altaffiltext{2}{Leiden Observatory, Leiden University, 
P.O. Box 9513, 2300-RA Leiden, The Netherlands}

\altaffiltext{3}{ISDEFE, Beatriz de Bobadilla 3, 28040 Madrid, Spain}

\altaffiltext{4}{Instituto de Astronom\'\i a, UNAM, Apartado Postal 70-264, CDMX C.P. 04510, Mexico}

\begin{abstract}

 We carried out multiwavelength (0.7-5 cm), multiepoch (1994-2015) Very 
Large Array (VLA) observations toward the region enclosing the bright 
far-IR sources FIR 3 (HOPS 370) and FIR 4 (HOPS 108) in OMC-2. We report 
the detection of 10 radio sources, seven of them identified as young 
stellar objects. We image a well-collimated radio jet with a thermal 
free-free core (VLA 11) associated with the Class I intermediate-mass 
protostar HOPS 370. The jet presents several knots (VLA 12N, 12C, 12S) 
of non-thermal radio emission (likely synchrotron from shock-accelerated 
relativistic electrons) at distances of $\sim$ 7,500-12,500 au from the 
protostar, in a region where other shock tracers have been previously 
identified. These knots are moving away from the HOPS 370 protostar at 
$\sim$ 100 km s$^{-1}$. The Class 0 protostar HOPS 108, which itself is 
detected as an independent, kinematically decoupled radio source, falls 
in the path of these non-thermal radio knots. These results favor the 
previously proposed scenario where the formation of HOPS 108 has been 
triggered by the impact of the HOPS 370 outflow with a dense clump. 
However, HOPS 108 presents a large proper motion velocity of $\sim 
$30 km s$^{-1}$, similar to that of other runaway stars in Orion, whose origin would be puzzling within this scenario. Alternatively, an apparent proper motion could result because of changes in the position of the centroid of the source due to blending with nearby extended emission, variations in the source shape, and/or opacity effects.

\end{abstract}



\section{Introduction} \label{sec:intro}

OMC-2 is an active star-forming region (e.g., \citealt{PetMeg2008}) in 
the Orion A molecular cloud, located at a distance of 414 $\pm$7 pc 
\citep{Menten2007,Kim2008,Kounkel2017}. \citet{Mezger1990} identified 
six bright mm/IR sources (FIR 1-6) within a region of about $6'$ in size 
that have been associated with young stellar objects (YSOs) through 
subsequent studies (\citealt{Adams2012}, \citealt{Furlan2014,Furlan2016} 
and references therein). The region has been imaged at mm and submm 
wavelengths by \citet{Chini1997} and \citet{Lis1998}, and in the near- 
and mid-IR by \citet{Tsujimoto2003}, \citet{Nielbock2003}, and 
\citet{Megeath2012}. At mm and submm wavelengths the brightest source is 
FIR 4, which has been associated with the HOPS 108 Class 0 protostar 
(\citealt{Adams2012}, \citealt{Furlan2016})\footnote{In low resolution 
observations FIR 4 probably includes emission from other nearby 
objects.}. This source is connected through a filamentary cloud 
structure to the bright source FIR 3, also known as HOPS 370, an 
intermediate-mass Class I YSO with an $L_{\rm bol} \sim 360~L_{\odot}$ 
(\citealt{Adams2012}, \citealt{Furlan2016}) located about $30''$ to the 
NE (see Fig. \ref{fig:2014} for the positions and nomenclature of the 
sources).  The region was observed with the Very Large Array (VLA) at 
3.6 cm in the D configuration (angular resolution $\sim8''$) by 
\citet{Reipurth1999}. These authors found a weak elongated source (VLA 
12) toward FIR 4 (HOPS 108) that was interpreted as a radio jet 
originating from this protostar. A stronger, but unresolved source (VLA 
11), was detected toward FIR 3 (HOPS 370).

The nature of the source FIR 4 is still uncertain. \citet{Shimajiri2008} 
through mm and submm observations proposed that FIR 4 was composed of 11 
dusty cores. These authors detected molecular line emission from shock 
tracers (SiO and CH$_3$OH) in the proximity of FIR 4 and proposed that 
the interaction of a powerful molecular outflow driven by FIR 3 (e.g., 
\citealt{Williams2003}, \citealt{Takahashi2008}) with a dense clump 
could be triggering the formation of a next generation of stars at the 
position of FIR 4. On the other hand, \citet{LopSep2013} based on 2 mm 
observations proposed that FIR 4 was composed of at least three cores, 
and suggested that its associated 3.6 cm source VLA 12 traces an HII 
region photoionized by an early-type (B3-B4) star with a luminosity of 
700-1000~$L_{\odot}$ embedded within one of these cores. A bolometric 
luminosity $L_{\rm bol}\simeq1000~L_{\odot}$ for FIR 4 was obtained by 
\citet{Crimier2009} but this estimate, based on low angular resolution 
IRAS data, is very uncertain because of the difficulties in separating 
the emission of the protostar from that of the surrounding molecular 
cloud and other neighboring objects.

In a rigorous analysis of the available photometry of HOPS 108, 
including Herschel measurements that constrain better the peak of the 
Spectral Energy Distribution (SED) than the IRAS fluxes, 
\citet{Furlan2014} give $L_{\rm bol} = 37~L_{\odot}$. This is much lower 
than the earlier estimate by \citet{Crimier2009} or the value adopted by 
\citet{LopSep2013}. Due to uncertainties in the amount of external 
heating and the inclination of the protostar, \citet{Furlan2014} find 
that they can fit the SED with models having values for the intrinsic 
total luminosity\footnote{The intrinsic total luminosity of a source and 
its bolometric luminosity derived from the observed flux densities can 
differ in case of an inhomogeneous surrounding medium (e.g., if outflow 
cavities are present) resulting in an anisotropic radiation field where 
the observed flux density depends on inclination angle (see 
\citealt{Furlan2016}).}  ranging from 15~$L_{\odot}$ to 540~$L_{\odot}$, 
although models with luminosities $\le 100~L_{\odot}$ are favored.

\citet{Manoj2013} detected far-IR CO lines indicative of shock-heated 
gas at high temperatures ($\ga 2000$ K) toward both HOPS 108 and HOPS 
370. Recently, \citet{GonGar2016} imaged in far-IR [OI] lines and in the 
submm CO(6-5) line, a powerful ($\dot 
M\simeq2\times10^{-6}~M_\odot$~yr$^{-1}$, as estimated by these authors) 
bipolar jet/outflow originating from HOPS 370. The outflow extends to 
the NE of HOPS 370 (FIR 3) in the direction where a 4.5 $\mu$m 
Spitzer/IRAC image (which traces shock-excited H$_2$ emission) shows an 
extended jet that terminates in a bow shock \citep{Megeath2012}. To the 
SW, the outflow terminates near the location of HOPS 108, with a bright 
[OI] emission peak that \citet{GonGar2016} argue originates in the 
terminal shock produced by the jet.

In summary, \citet{Shimajiri2008}, \citet{LopSep2013}, and 
\citet{GonGar2016} suggest that the protostar HOPS 108 results from the 
interaction of an outflow driven by HOPS 370 with a dense clump. Within 
this scenario, \citet{LopSep2013} assume that a star cluster is being 
formed, with the radio source VLA 12 tracing an HII region associated with 
the most massive star; for this star they adopt a luminosity of $\sim$1000 
$L_\odot$, similar to earlier estimates of \citet{Crimier2009}. On the 
other hand, \citet{Furlan2014}, \citet{Reipurth1999}, and 
\citet{Adams2012} interpret VLA 12 as a radio jet driven by a modest 
luminosity protostar associated with HOPS 108.

In this paper we analyze new and archive VLA observations that shed new 
light on the origin of the HOPS 108 protostar, the nature of its radio 
emission, and its relationship with other nearby sources.

\section{Observations} \label{sec:observations}

The observations were carried out with the Karl G. Jansky VLA of the 
National Radio Astronomy Observatory (NRAO)\footnote{NRAO is a facility 
of the National Science Foundation operated under cooperative agreement 
by Associated Universities, Inc.} in the C configuration in 2014 
(Project 14B-296) at C (5 cm), X (3 cm), K (1.3 cm) and Q (0.7 cm) 
bands, and in the A configuration in 2015 (Project 15A-369) at C-band (5 
cm). In both cases, the phase center was close to the position of HOPS 
108.  Further details of the observational setup are given in Table 
\ref{tab:observations}. The data were edited and calibrated using the 
Common Astronomy Software Applications (CASA; version 4.2.2) package.

We also analyzed several epochs of VLA archive data at X-band (3.6 cm) 
obtained with different configurations between 1994 and 2000 (see Table 
\ref{tab:observations} for details). These data were edited and 
calibrated using the Astronomical Image Processing System (AIPS), and 
then were concatenated using the CASA package.

All the images were made with the CASA task CLEAN using multi-frequency 
synthesis \citep{Conway1990} and fitting the frequency dependence of the 
emission with a Taylor series expansion with nterms = 2 during the 
deconvolution. There is extended emission of the bright HII region M43 
(NGC 1982) $\sim5'$ to the south of HOPS 108 that makes it difficult the 
imaging of the archive (X-band) data and the new data at C-band. In 
order to do a proper cleaning of our target sources we downweighted the 
extended emission by removing the shortest baselines. For the archive 
and the 2014 C-band data we used only baselines $>$5 k$\lambda$ and the 
multiscale deconvolution algorithm to make the images. For the 2015 
A-configuration data, which are more sensitive, we used only baselines 
$>$25 k$\lambda$.

\section{Results and Discussion} \label{sec:results}

In our new data of epochs 2014 and 2015 we detect the radio sources VLA 
11, VLA 12, and VLA 13 previously reported at 3.6 cm by 
\citet{Reipurth1999}, but our observations cover additional 
bands and are more sensitive, revealing further details. The FWHM of our 
primary beam, centered near VLA 12, ranges from $\sim1'$ at Q-band to 
$\sim8'$ at C-band. Therefore, we detected additional sources, including 
several of the IR sources in the field. In this paper we will discuss 
only the sources detected in the proximity (within $\la 30''$) of HOPS 
108. Positions of these sources are given in Table \ref{tab:positions}. 
The remaining detected sources will be discussed in a forthcoming 
paper (A. K. D\'\i az-Rodr\'\i guez et al. 2017, in preparation).

\subsection{Identification of the Detected Sources} \label{identification}

\begin{figure}[h!]
\figurenum{1}
\includegraphics[width=0.49\textwidth]{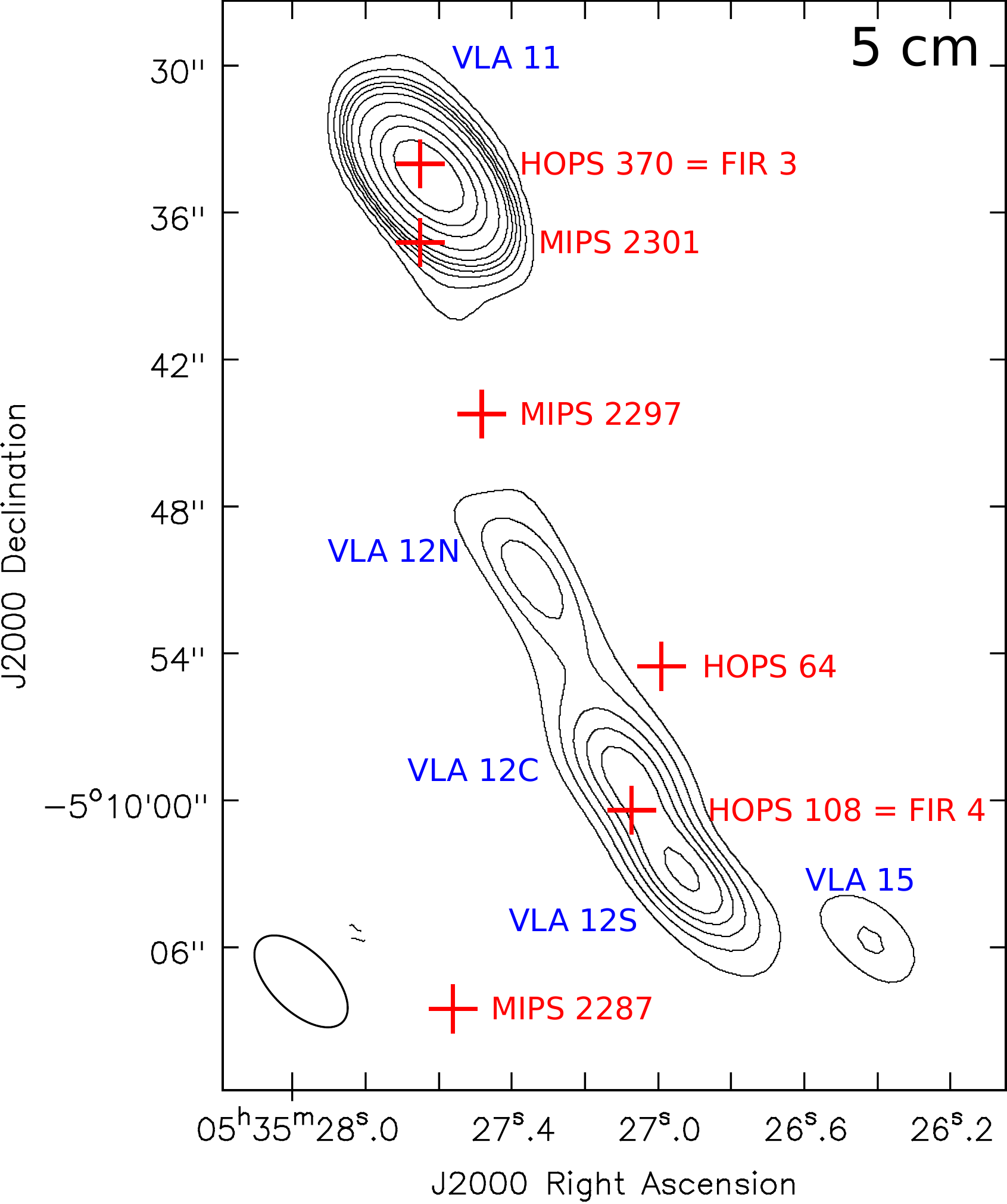} 
\hfill
\includegraphics[width=0.49\textwidth]{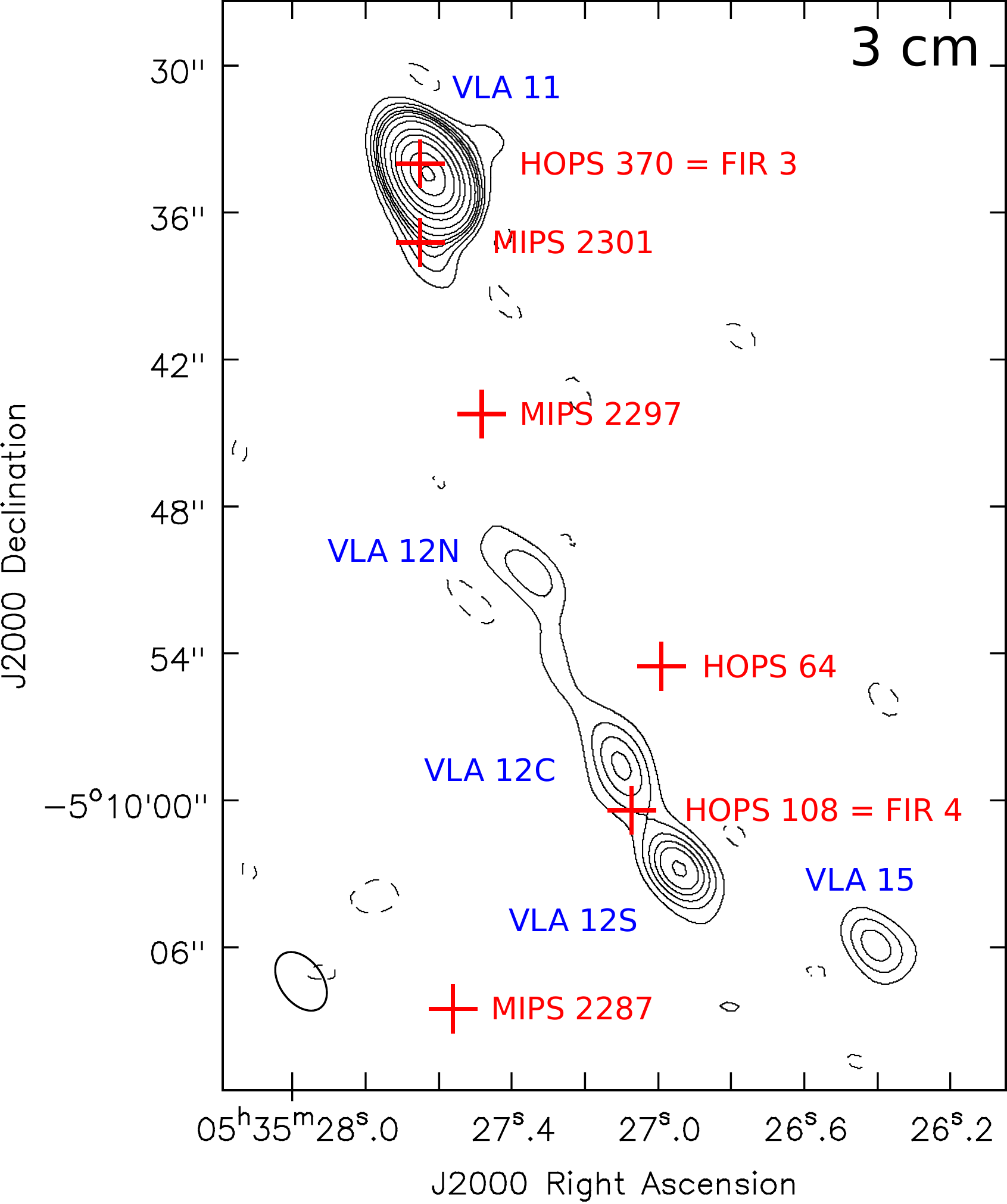}
\vspace{0.3cm}
\includegraphics[width=0.49\textwidth]{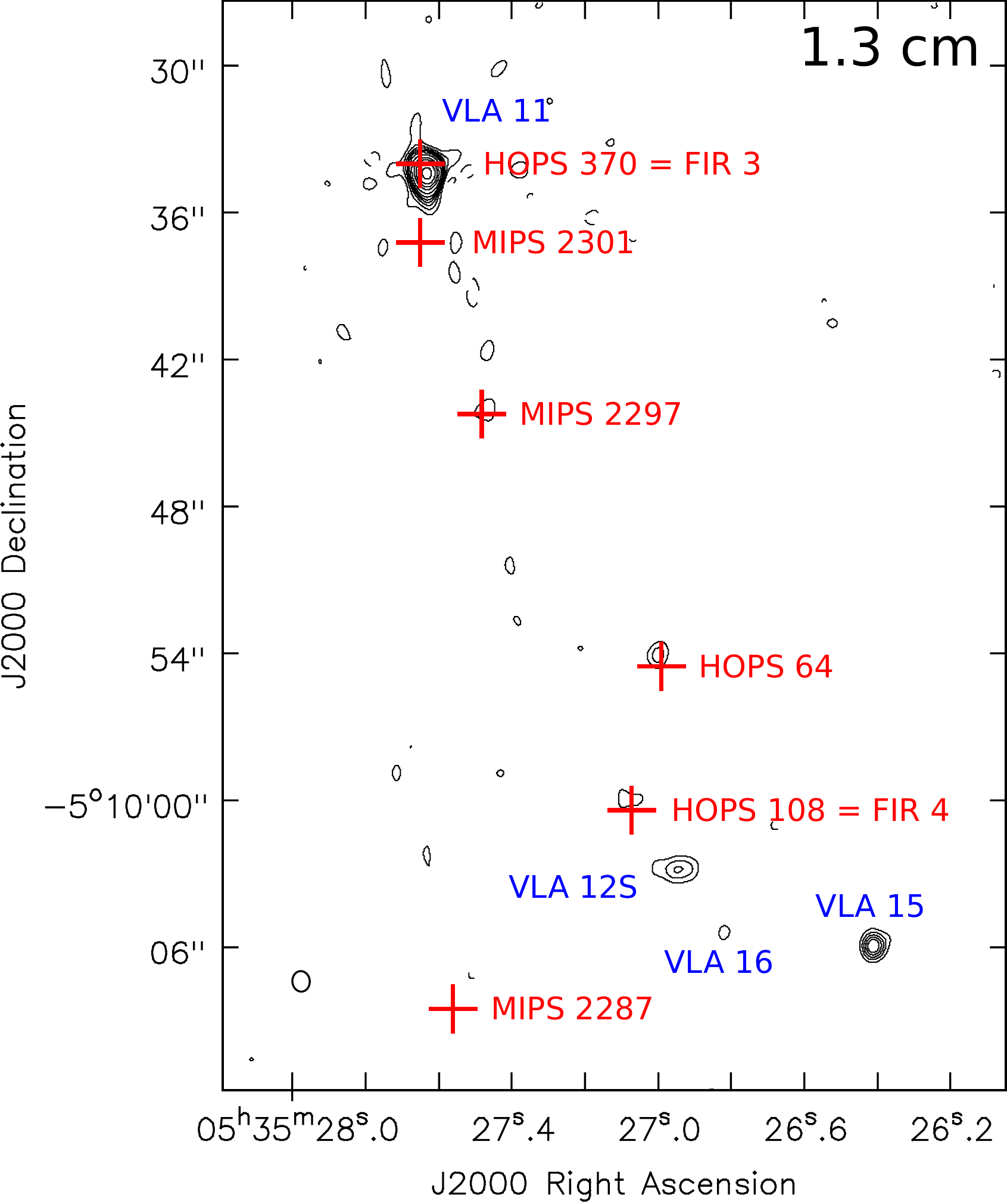} 
\hfill
\includegraphics[width=0.49\textwidth]{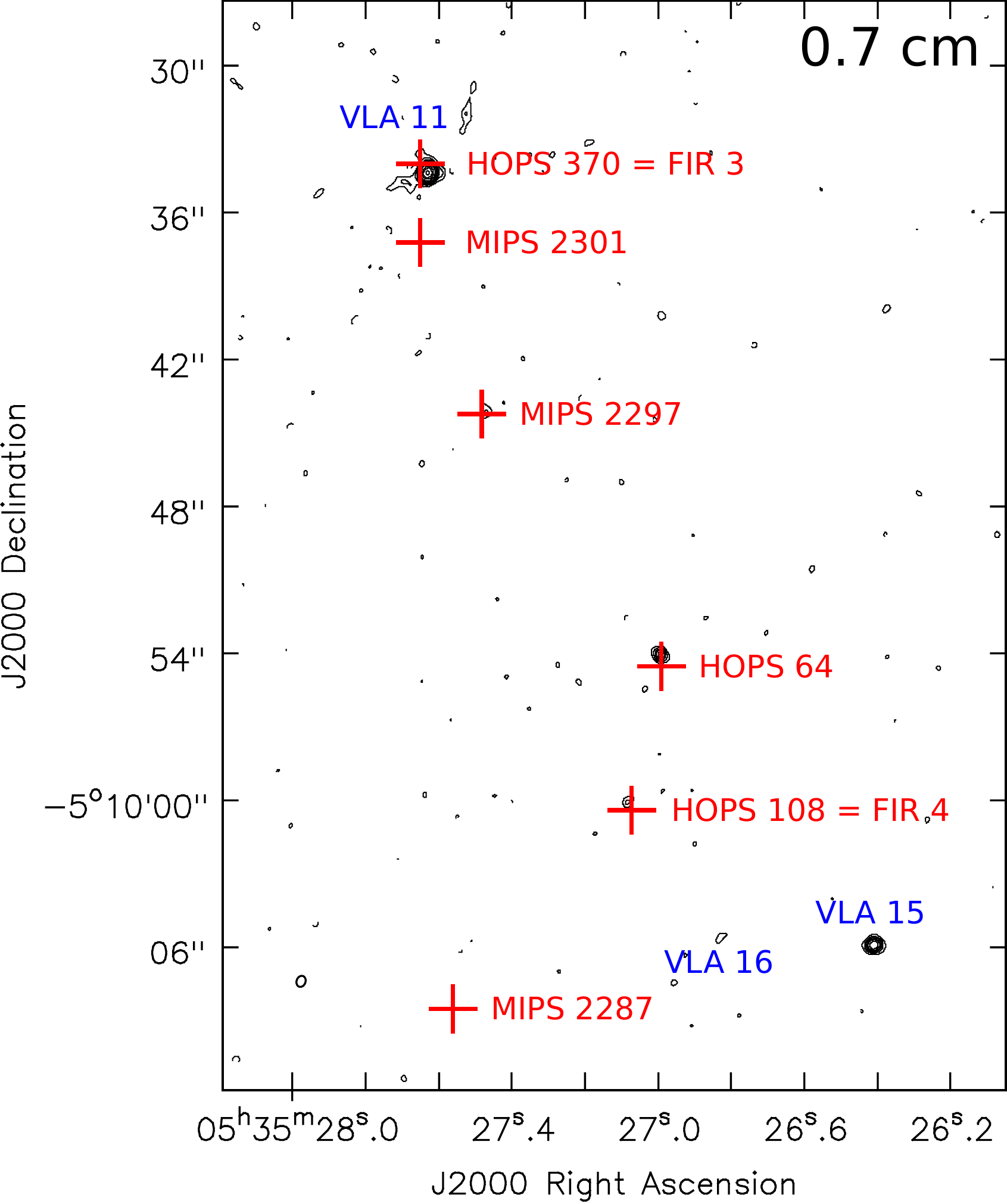}
 \caption{\scriptsize VLA C-configuration maps (epoch 2014) of the 
region around the HOPS 108 protostar. Contour levels are $-$3, 3, 6, 9, 
12, 15, 20, 30, 50, 70, 100, 150 and 200 times the rms noise of each 
map. The robust weighting parameter \citep{Briggs1995} was set to 0. The 
synthesized beam is shown at the bottom left corner of each map. The 
10.4 {\micron} positions (\citealt{Nielbock2003}) are indicated by red 
plus signs whose size corresponds to the positional uncertainty. For 
HOPS 108, that is not reported at 10.4 {\micron}, the 24 {\micron} 
position \citep{Megeath2012} is given.
 Top left: C band (5 cm); rms = 11 $\mu$Jy beam$^{-1}$; HPBW = 
4\farcs75$\times$2\farcs46, P.A. = 46\degr. Top right: X band (3 cm); 
rms = 9 $\mu$Jy beam$^{-1}$; HPBW = 2\farcs68$\times$1\farcs73, P.A. = 
36\degr. Bottom left: K band (1.3 cm); rms = 10 $\mu$Jy beam$^{-1}$; 
HPBW = 0\farcs84$\times$0\farcs71, P.A. = 5\degr. Bottom right: Q band 
(0.7 cm) map; rms = 15 $\mu$Jy beam$^{-1}$; HPBW = 
0\farcs46$\times$0\farcs36, P.A. =$-$20\degr.  \label{fig:2014}}
 \end{figure}

\begin{figure}[h!]
\figurenum{2}
\includegraphics[width=0.47\textwidth]{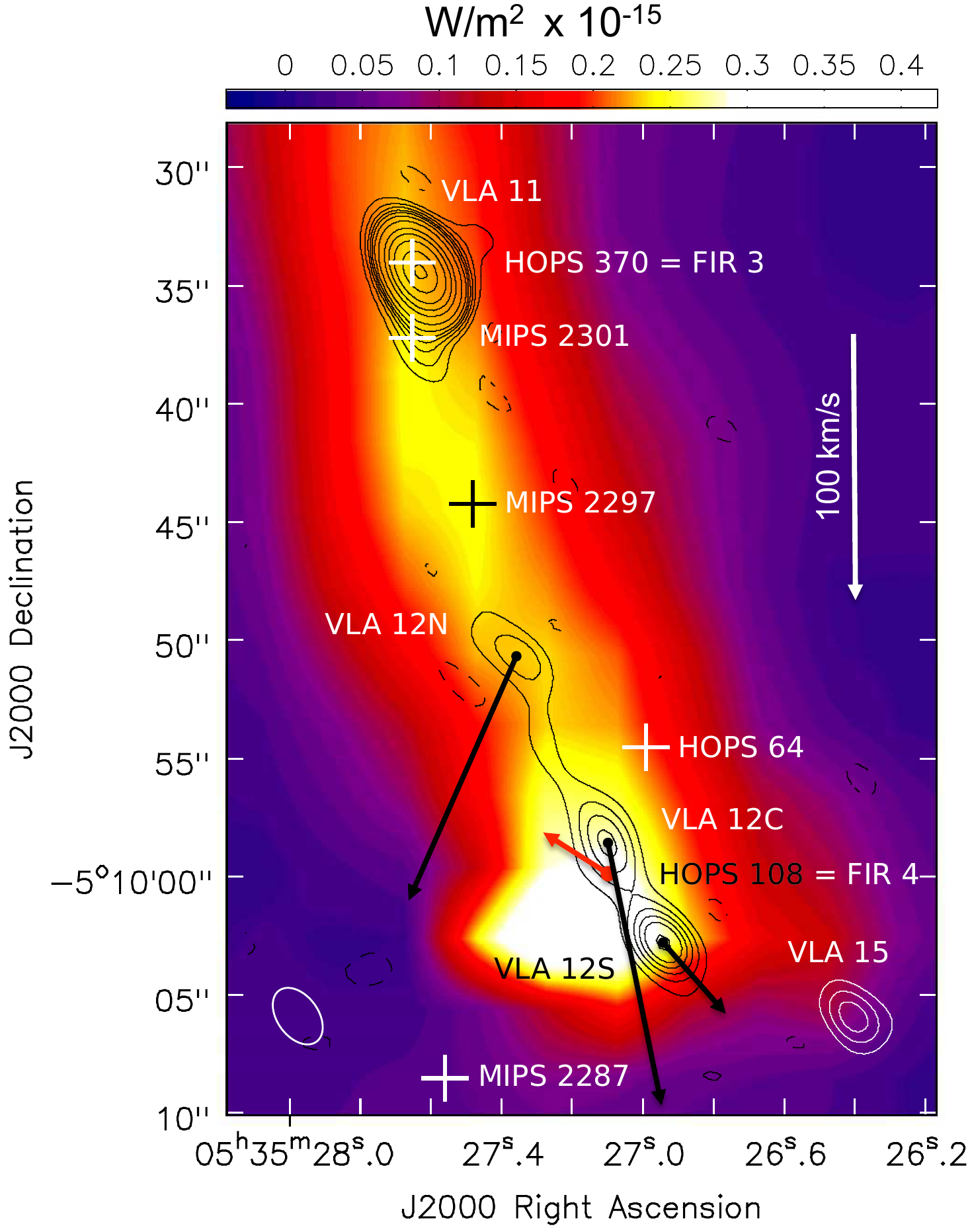} 
\includegraphics[width=0.53\textwidth]{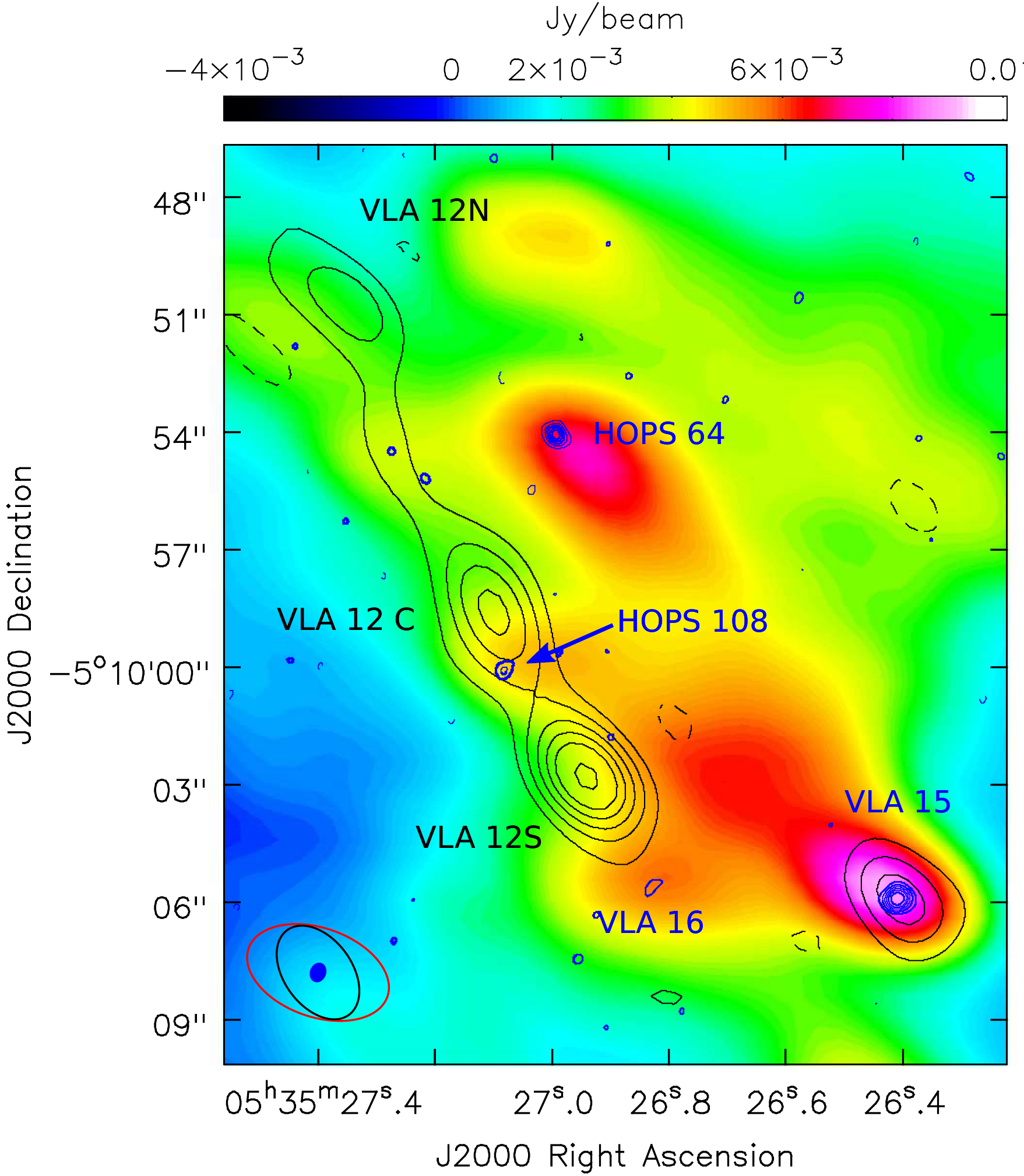}
 \caption{\scriptsize Left panel: Image of the [OI] 63 $\mu$m line 
emission (color scale, \citealt{GonGar2016}) overlaid on the 3 cm emission 
VLA map (contours, see Fig. \ref{fig:2014}) with proper motions 
indicated by arrows.  Both images are tracing the 
jet ejected from HOPS 370. Note that the [OI] 63 $\mu$m emission peaks 
close to the position of HOPS 108, suggesting that this protostar has been 
formed in a region where a strong shock interaction between the jet and 
the ambient medium is taking place. Plus signs indicate the positions of IR sources.
Right panel: ALMA image of the dust 
emission at 3 mm (color scale, \citealt{Kainulainen2016}) overlaid on 
the 
VLA maps at 3 cm (black contours) and 0.7 cm (blue contours, see Fig. 
\ref{fig:2014}). Note that the radio jet at 3 cm appears to delineate the 
eastern edge of the dust clump traced by the ALMA image. 
\label{fig:overlays}}
\end{figure}

In Figure \ref{fig:2014} we show the maps of VLA 11 and VLA 12 at 
different wavelengths obtained in our 2014 VLA C-configuration 
observations. VLA 12 (in the proximity of FIR 4/HOPS 108) is clearly 
detected as a very elongated source at 5 and 3 cm, showing three 
knots that we call VLA 12N, VLA 12C, and VLA 12S, for the northern, 
central and southern observed emission peaks, respectively. At first 
glance, these results appear to confirm VLA 12 as a radio jet, as was 
first suggested by \citet{Reipurth1999}. These knots coincide with the 
southern, brightest part of the jet/outflow traced by the far-IR [OI] 
and submm CO(6-5) lines, which is assumed to originate in HOPS 370 
(\citealt{GonGar2016}; see Fig.~\ref{fig:overlays}, left). The IR 
position of the HOPS 108 protostar (\citealt{Megeath2012}) falls close 
to (within $\sim2''$) the position of knot VLA 12C, apparently favoring 
HOPS 108 as the driving source of the proposed VLA 12 radio jet.

\begin{figure}[h!]
\figurenum{3}
\includegraphics[scale=0.9]{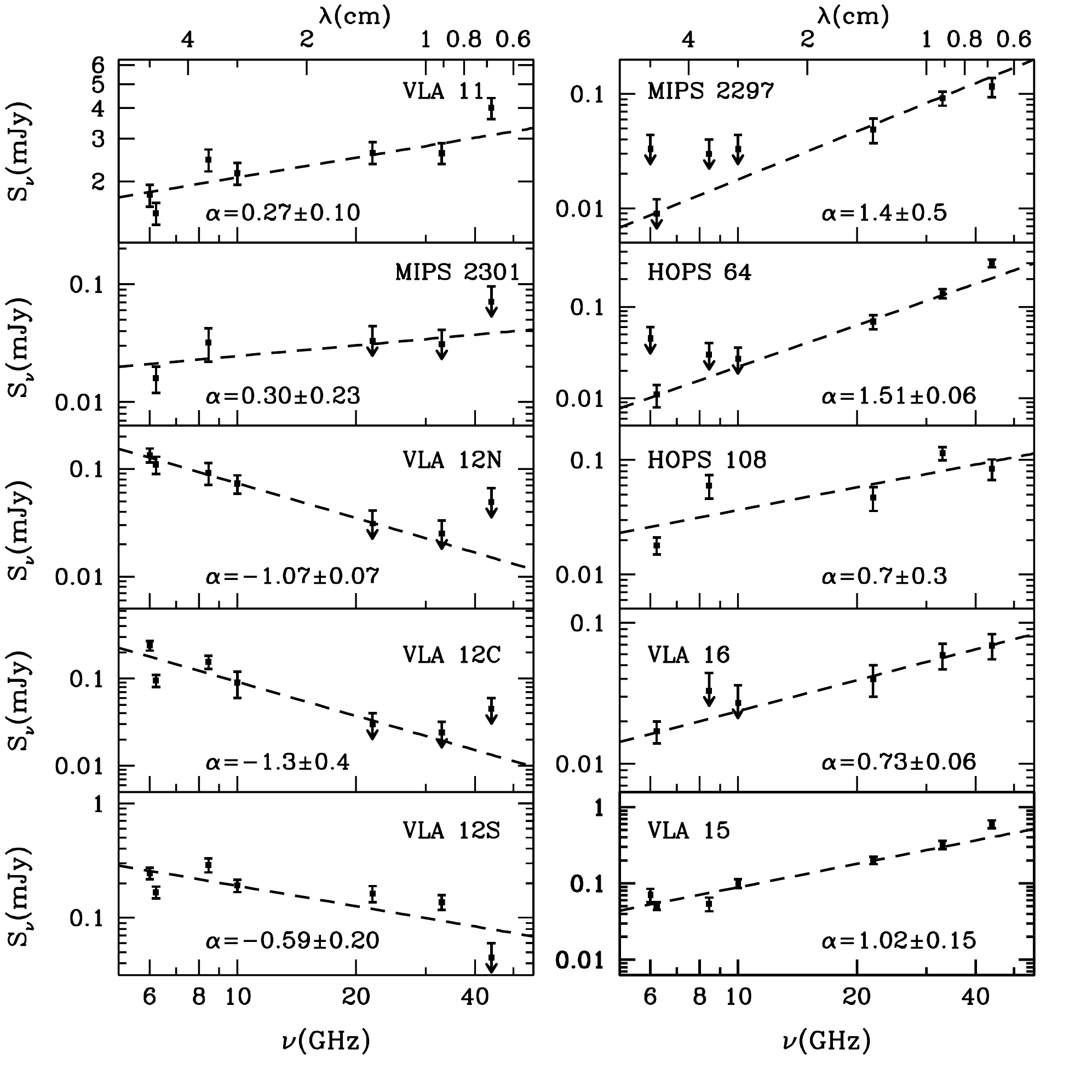}
\caption{\scriptsize Spectra of the detected VLA sources. Error bars are 1\,$\sigma$. 
Upper limits are represented by arrows where the central symbol is 
at 3\,$\sigma$. Dashed lines represent least-squared fits. To avoid 
possible dust contamination the 0.7 cm datapoint has been excluded in 
the fitting for the positive spectral index sources.  Upper limits have 
been taken into account only when they constrain the fit.
\label{fig:SED}}
\end{figure}

At shorter wavelengths (1.3 and 0.7 cm; bottom panels in Fig. 
\ref{fig:2014}) the emission of the VLA 12 knots decreases making them 
undetectable, indicating that they have a negative spectral index 
$\alpha$ (where $S_\nu \propto \nu^\alpha$), characteristic of 
non-thermal emission. This is confirmed by the analysis of the spectra 
obtained from our full dataset (Table \ref{tab:pmotions}, Fig. 
\ref{fig:SED}). Despite the possible time variability both in flux 
density (that can be more prominent in non-thermal emission) and 
morphology, as well as the difficulties inherent in the measure of 
extended emission from data taken with different angular resolutions, 
the obtained spectral indices in the centimeter range are negative for 
the three knots of VLA 12, with values of $-$1.1, $-$1.3, and $-$0.6. 
The behavior is clear even if only the C-configuration data from October 
2014, which were taken almost simultaneously, are considered. Also, we 
have checked in these data that the negative spectral indices obtained 
are not spurious results of differences in the uv coverage and/or 
angular resolution, that could make data at shorter wavelengths less 
sensitive to extended emission. To do that, we obtained pairs of images 
of contiguous bands using only baselines in the same uvrange. Using 
these pairs of maps we confirmed that the flux density truly decreases 
at shorter wavelengths, resulting in negative spectral indices for the 
VLA 12 knots. These negative spectral indices cannot be explained by 
thermal emission (see \citealt{Rodriguez1993}) and should be attributed 
to non-thermal emission, probably arising from relativistic electrons 
accelerated in strong shocks (\citealt{CarGon2010}; see below). This 
rules out the HII region scenario proposed by \citet{LopSep2013} for VLA 
12.

\begin{figure}[h!]
\figurenum{4}
\begin{center}
\includegraphics[width=0.38\textwidth]{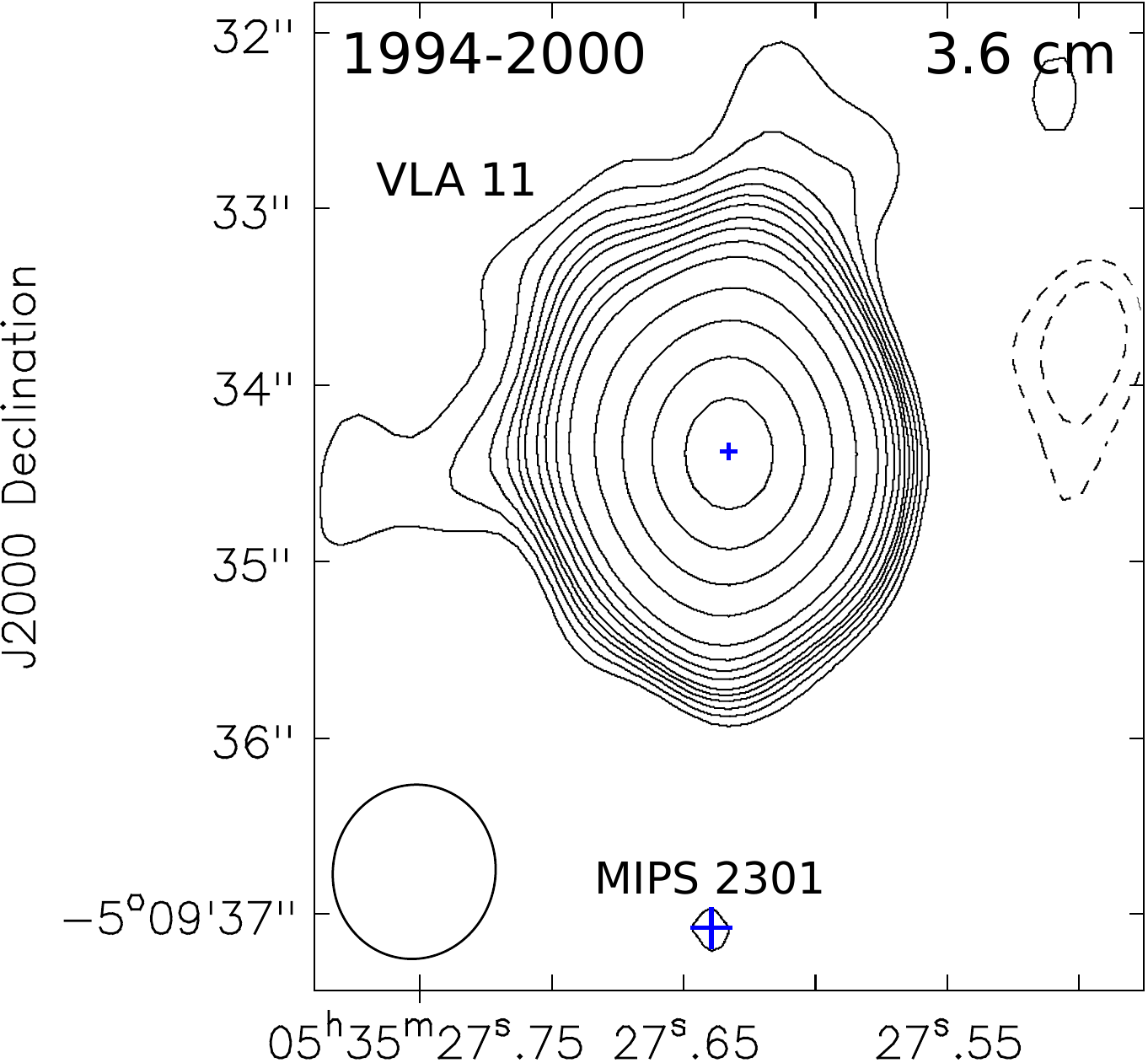} 
\hspace{1.6cm}
\vspace{0.2cm}
\includegraphics[width=0.38\textwidth]{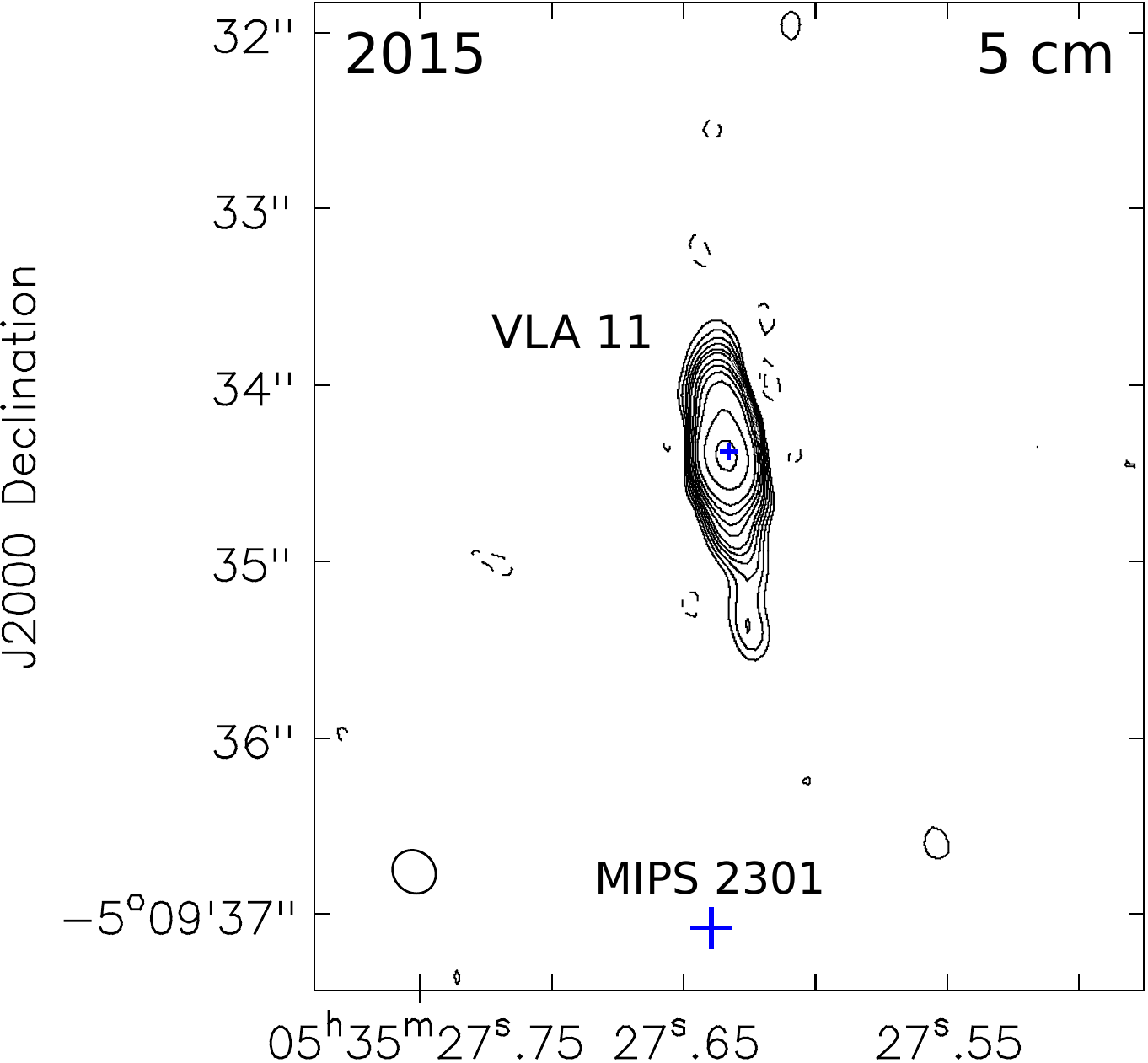} 
\vspace{0.2cm}
\includegraphics[width=0.38\textwidth]{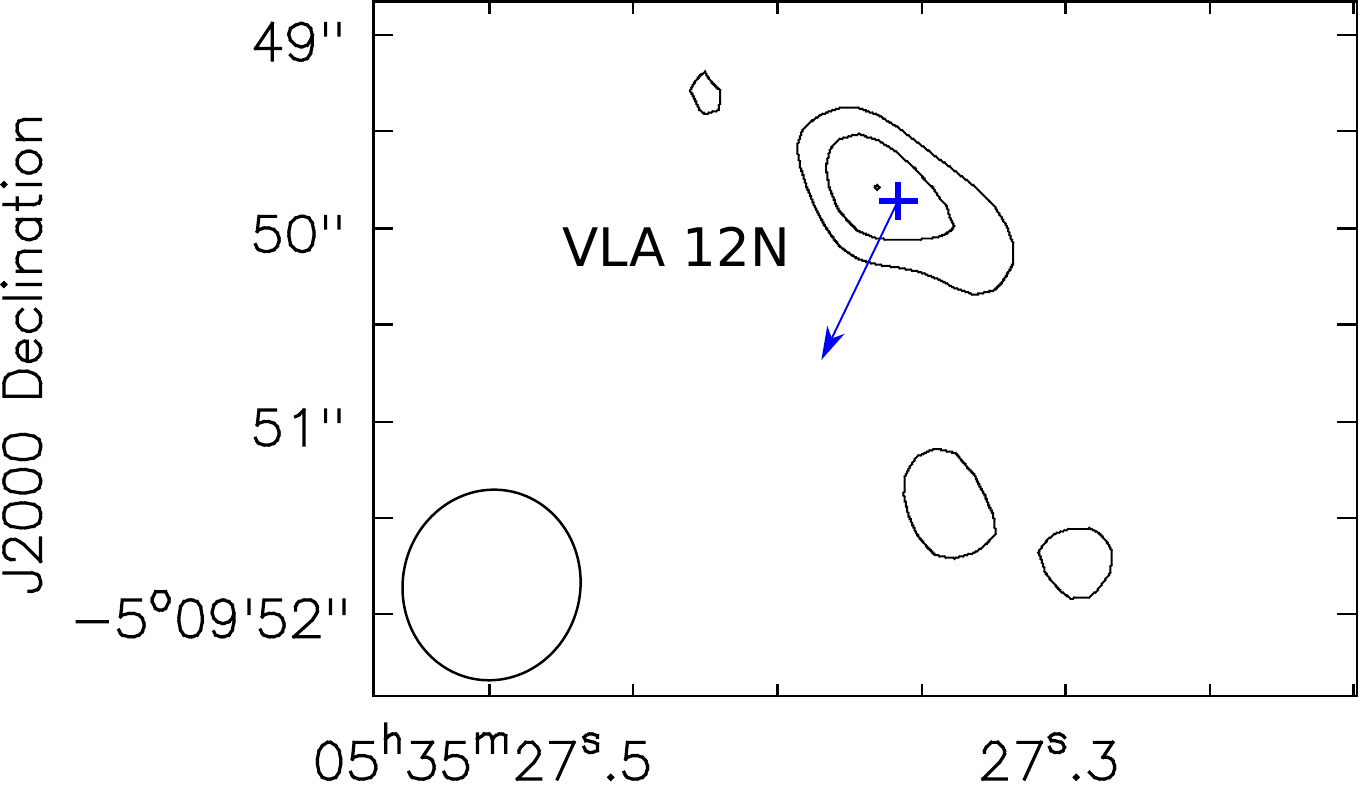} 
\hspace{1.6cm}
\includegraphics[width=0.38\textwidth]{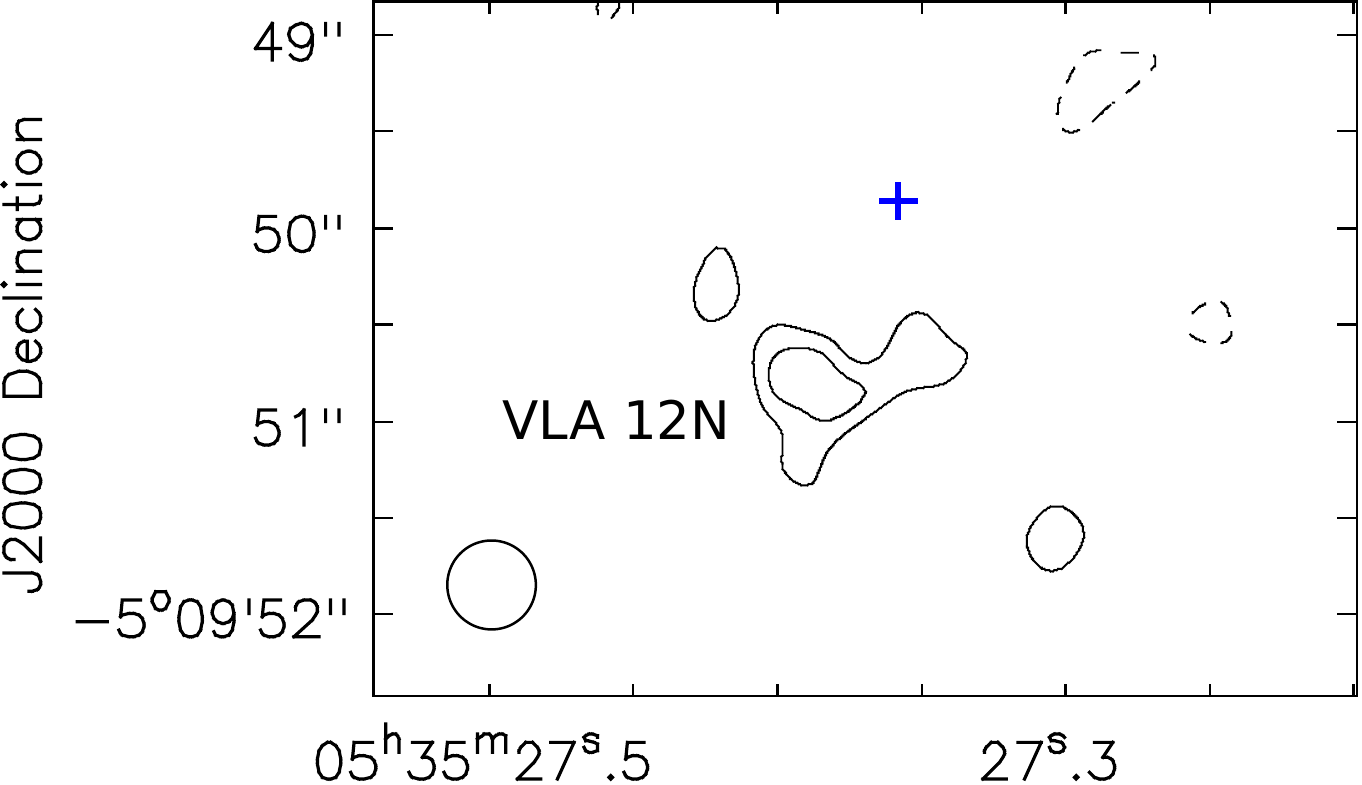} 
\vspace{0.2cm}
\includegraphics[width=0.38\textwidth]{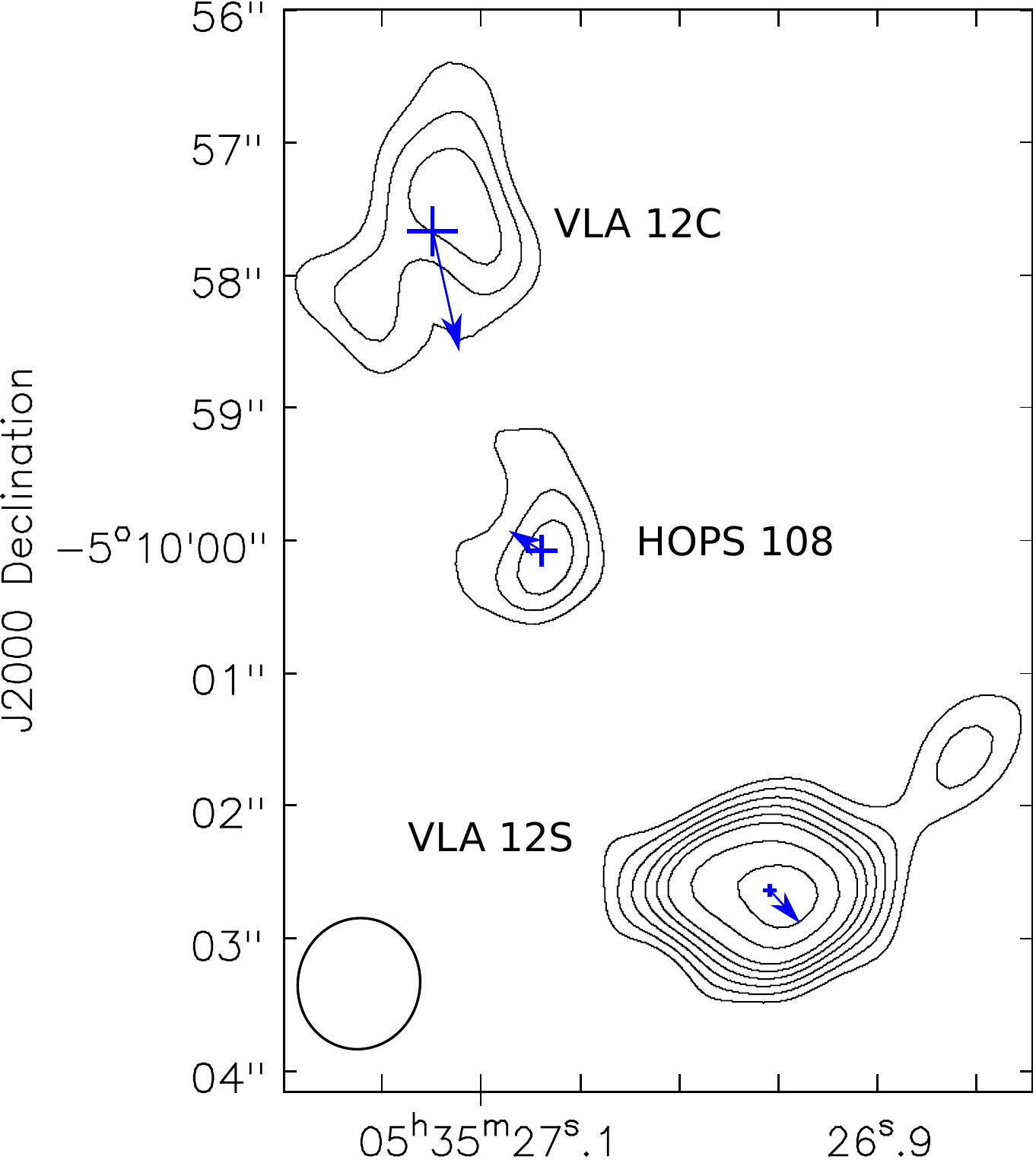} 
\hspace{1.6cm}
\includegraphics[width=0.38\textwidth]{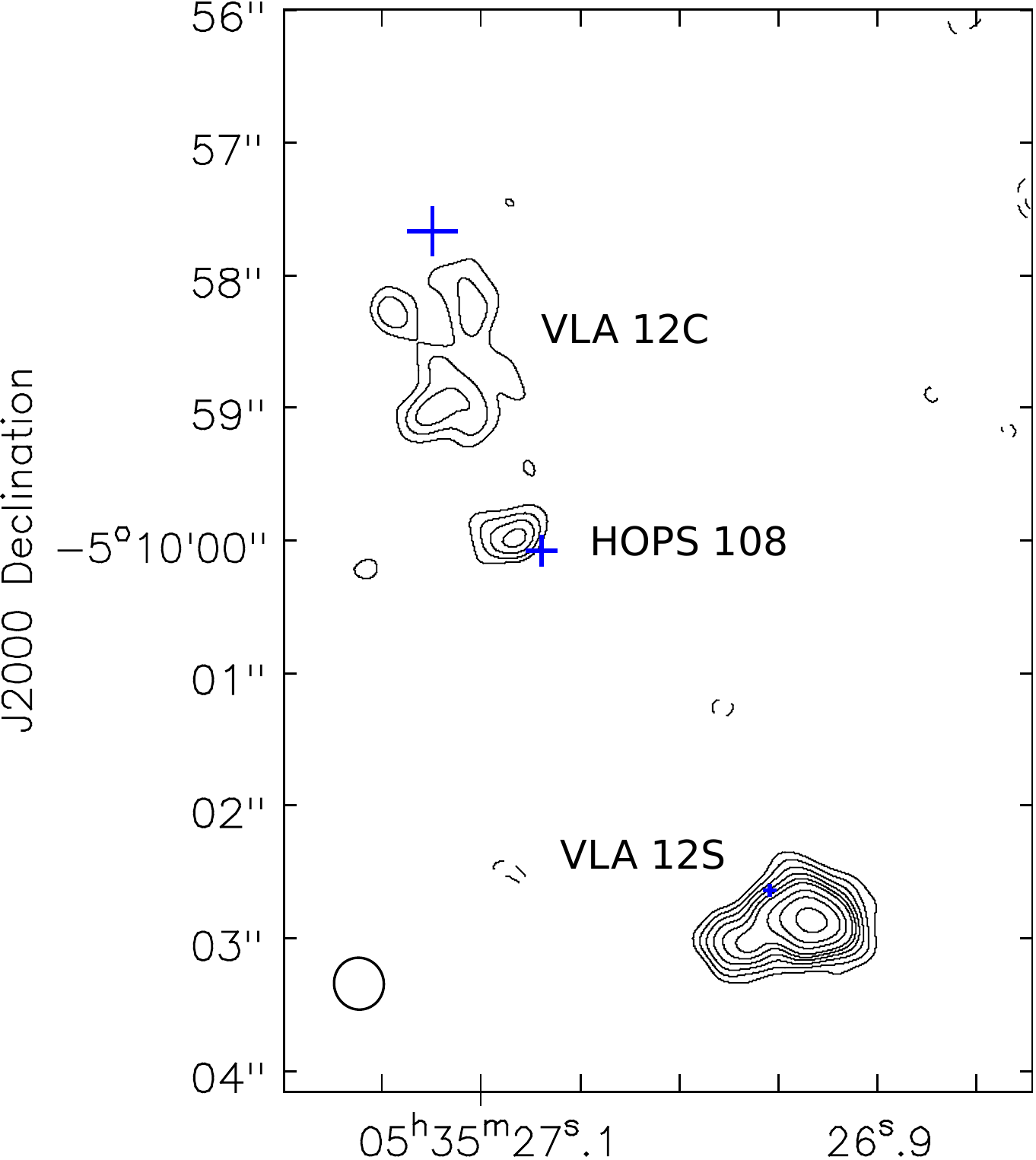}
\vspace{0.2cm}
\includegraphics[width=0.38\textwidth]{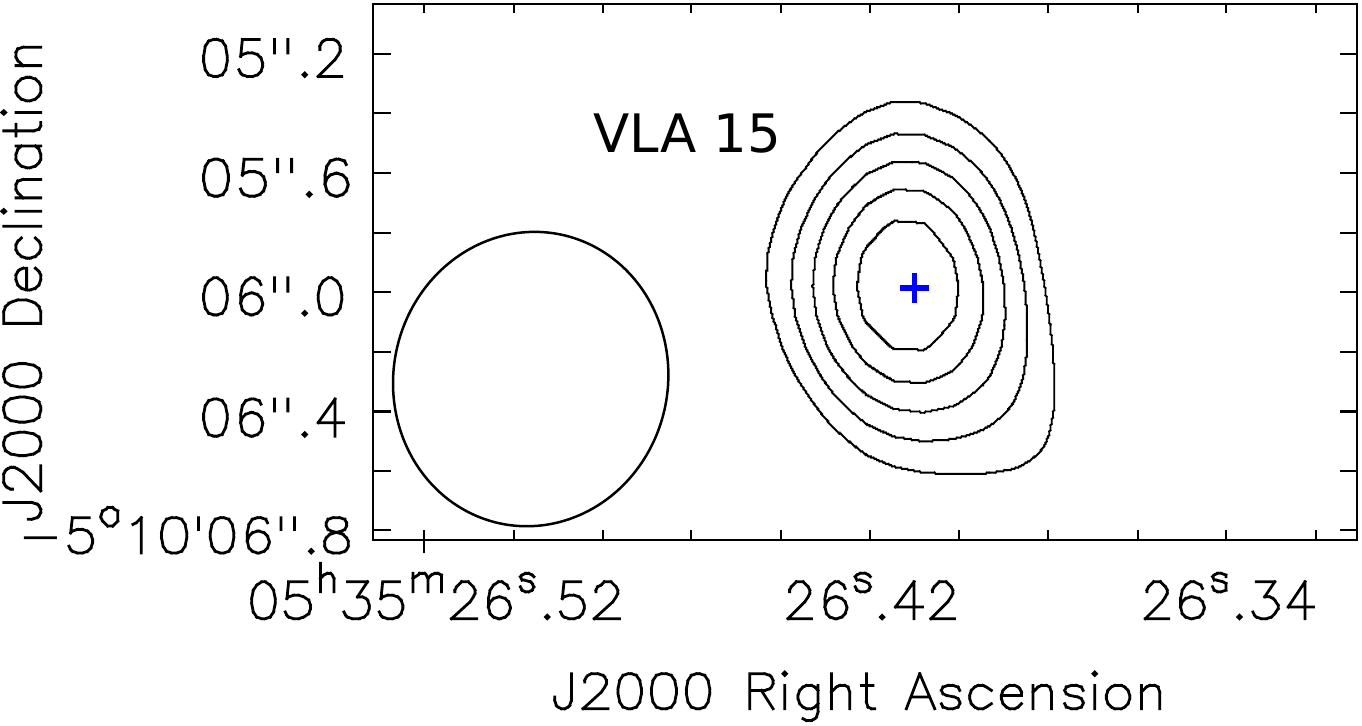} 
\hspace{1.6cm}
\includegraphics[width=0.38\textwidth]{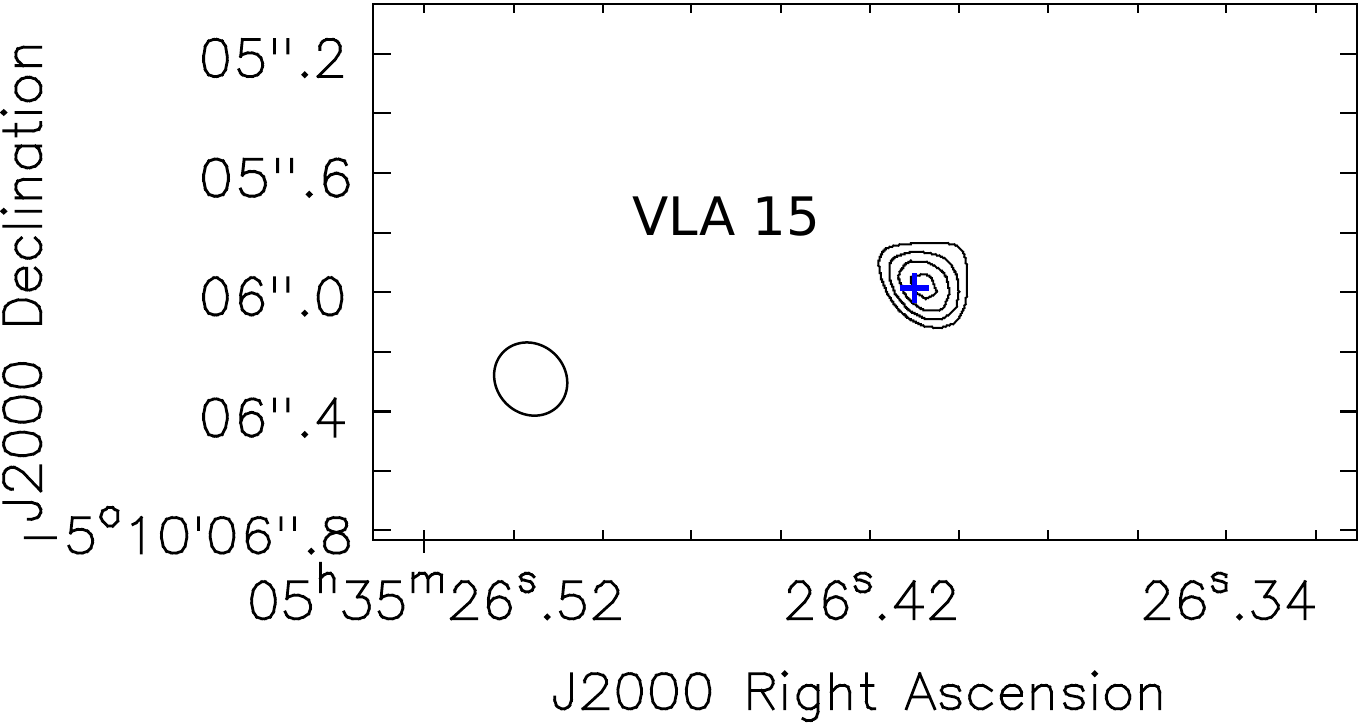} 
\end{center}
 \caption{\scriptsize Proper motions of VLA sources. First epoch maps 
(X-band; left panels) were obtained by combining archive data from 1994 
to 2000 (Table \ref{tab:observations}). Second epoch maps (C-band; right 
panels) were obtained from the 2015 data. Different weightings have been 
used to better emphasize the emission of each source. Synthesized beams 
of $\sim0\farcs95$ and rms of $\sim$ 10 $\mu$Jy beam$^{-1}$ were 
obtained for the first epoch maps, while synthesized beams of 
$\sim0\farcs24$-$0\farcs46$, and rms of 3-8 $\mu$Jy beam$^{-1}$ were 
obtained for the second epoch maps. Contour levels are $-$4, $-$3, 3, 4, 
5, 6, 7, 8, 10, 12, 15, 20, 35, 60, 100, and 150 times the rms noise of 
each map. Positions of the sources derived from elliptical Gaussian fits 
to the first epoch maps are marked with blue plus signs, whose sizes 
represent their positional errors. Arrows indicate the displacement 
between the two epochs. \label{fig:movprop}}
 \end{figure}

Although the position of the knot VLA 12C at 6 and 3 cm appears somewhat 
($\sim1.5''$) displaced to the NE of the position of HOPS 108 (as 
obtained from 24 and 8 {\micron} data; \citealt{Megeath2012}), at 1.3 
and 0.7 cm the radio emission appears to be closer to the HOPS 108 IR 
position (see Fig.~\ref{fig:2014}). This shift in the position could 
indicate either an opacity gradient or that the radio emission observed 
at shorter wavelengths originates in a different object than the 
emission observed at longer wavelengths. This last suggestion is 
confirmed by our data of subarcsecond angular resolution at long 
wavelengths (Fig.~\ref{fig:movprop}) that reveal two radio sources 
spatially separated by $\sim2''$. We identify the northeasternmost of 
these two radio sources with the VLA 12C knot and the other source, 
whose position coincides with the IR source, with the radio counterpart 
of the HOPS 108 protostar. Recent ALMA observations with an angular 
resolution of 0\farcs12 (J. Tobin et al. 2017b, in preparation) reveal a 
compact source at 870 $\mu$m whose position coincides within $<$ 
0\farcs1 with the cm position (Table \ref{tab:pmotions}) confirming the 
nature of HOPS 108 as a protostar and its association with the cm radio 
source. Furthermore, the spectral index of HOPS 108 obtained from the cm 
data that separate its emission from that of VLA 12C shows that HOPS 108 
has a positive spectral index ($\alpha=0.7 \pm 0.3$; 
Table~\ref{tab:pmotions} and Fig.~\ref{fig:SED}) and, thus it is 
associated with thermal (free-free) emission, in contrast to the knots 
of VLA 12, whose emission is non-thermal.  From the empirical 
correlation between cm flux density and bolometric luminosity for 
protostellar objects, $(S_\nu\,d^2/{\rm mJy\,kpc^2})=0.008\,(L_{\rm 
bol}/L_\odot)^{0.6}$ \citep{Anglada2015}, and using our measured flux 
density at 3 cm (Table \ref{tab:pmotions}), we estimate a bolometric 
luminosity $L_{\rm bol}\simeq7~L_\odot$ for HOPS 108. This value of the 
bolometric luminosity is close to the lowest end of the range considered 
by \citet{Furlan2016} in the modeling of this object, being much smaller 
than early estimates ($L_{\rm bol}\simeq1000~L_\odot$; 
\citealt{Crimier2009,LopSep2013}) that were based on very low angular 
resolution data, and significantly smaller than the value obtained if 
the contribution of the VLA 12 knots is not removed ($L_{\rm 
bol}\simeq100~L_\odot$; \citealt{Furlan2014}). Thus, separating the 
emission of HOPS 108 from that of the VLA 12 knots confirms HOPS 108 as 
a low-luminosity object.

The radio source VLA 11, that was associated with FIR 3 (HOPS 370) and 
reported as unresolved by \citet{Reipurth1999}, is clearly resolved as a 
very elongated bipolar radio jet along a PA $\simeq5^\circ$ in our 
A-configuration data at 5 cm (Fig.~\ref{fig:movprop}). The radio source 
is also resolved in the 3.6 cm image obtained by combining archive data 
from 1994-2000. The spectral index is $\alpha=0.27\pm0.10$ (Table 
\ref{tab:pmotions}, Fig.~\ref{fig:SED}), which is a reasonable value for 
a thermal free-free radio jet \citep{Anglada1996}. HOPS 370 is known to 
be associated with an outflow extending along a PA similar to that of 
the radio jet (\citealt{GonGar2016}; see Fig.~\ref{fig:overlays}, left). 
The morphology of VLA 11 appears to change in the 0.7 cm image 
(Fig.~\ref{fig:2014}), that shows weak extended emission in a direction 
perpendicular to the jet, suggesting a noticeable dust contribution at 
this wavelength. To prevent dust contamination, the 0.7 cm datapoint has 
been excluded in the fits to determine the spectral index of this and 
the remaining thermal sources.

In the lower resolution images at 6 and 3 cm (Fig.~\ref{fig:2014}), VLA 
11 appears blended with the radio emission of MIPS 2301 
(\citealt{Megeath2012}, an IR source also known as MIR 22, 
\citealt{Nielbock2003}) located $\sim3''$ south of VLA 11. MIPS 2301 is 
well separated, but only marginally detected, in the higher resolution 
3.6 cm image shown in Figure \ref{fig:movprop} and in the 
A-configuration images at 5 cm obtained with the robust weighting 
parameter set to values $>$ 1 (not shown in 
Fig.~\ref{fig:movprop})\footnote{Note that MIPS 2301 is not aligned 
along the VLA 11 jet axis (PA = $5^\circ$). Thus, its radio emission can 
be distinguished from the knots of the radio jet in the high angular 
resolution images.}. With the data currently available we obtain a 
positive spectral index, $\alpha=0.30 \pm 0.23$ (Table 
\ref{tab:pmotions}, Fig.~\ref{fig:SED}), consistent with thermal 
free-free emission from a YSO. More sensitive observations are required 
for a better characterization of the nature of the radio emission of 
this object.

The source MIPS 2297 (\citealt{Megeath2012}, also known as MIR 23, 
\citealt{Nielbock2003}) is marginally detected at 1.3 cm and 0.7 cm 
(Fig.~\ref{fig:2014}), but clearly detected in the 0.9 cm observations 
of J. Tobin et al. (2017a, in preparation). It shows a positive spectral 
index ($\alpha=1.5\pm0.5$) in the cm regime (Table~\ref{tab:pmotions} 
and Fig.~\ref{fig:SED}), confirming its nature as a YSO. So far, MIPS 
2297 has not been studied in great detail, but it is classified as a 
young star with disk (Class II) by \cite{Megeath2012}.

We also detect HOPS 64 (Fig.~\ref{fig:2014}), a Class I object 
\citep{Furlan2016} that in the cm range presents a positive spectral 
index of $\alpha=1.51\pm0.06$ (Table \ref{tab:pmotions}, 
Fig.~\ref{fig:SED}), suggesting optically-thick free-free emission from 
a YSO. The position of the radio source falls in between the FIR 4d and 
FIR 4e cores imaged by \citet{Shimajiri2008}, and is close to, but 
slightly displaced from, one of the main emission peaks of the dust 
cloud imaged by ALMA at 3 mm (\citealt{Kainulainen2016}; see 
Fig.~\ref{fig:overlays}, right). Nevertheless, the position of the radio 
source coincides accurately (within $<$ 0\farcs1) with that of a compact 
870 $\mu$m source recently detected with ALMA (J. Tobin et al. 2017b, in 
preparation).

A few arcsec to the SW of the knot VLA 12S we detect a weak source. The 
source has not been reported previously at other wavelengths and we call 
it VLA 16. The detection is marginal in the individual bands ($\sim$ 
5\,$\sigma$ at most) but we consider that the source is real because it 
appears in four of the five observed bands (see Table 
\ref{tab:pmotions}). The spectral index is $\sim$ 0.7 
(Fig.~\ref{fig:SED}), consistent with that of a YSO/jet. The position of 
the radio source falls near a secondary emission peak in the 3 mm ALMA 
image \citep{Kainulainen2016} shown in Figure~\ref{fig:overlays} and 
coincides very well (within $<$ 0\farcs1) with the position of a weak 
870 $\mu$m source recently detected with ALMA observations of higher 
angular resolution (J. Tobin et al. 2017b, in preparation). From the 
empirical correlation between cm emission and bolometric luminosity 
\citep{Anglada2015} we infer $L_{\rm bol}\simeq1~L_\odot$. All these 
results confirm the reality of the source VLA 16 and its nature as an 
embedded low-luminosity YSO.

Finally, we report the detection in all the observed bands of a compact 
source, located $\sim8.5''$ SW of VLA 12S, and that we call VLA 15 
(Tables \ref{tab:positions}, \ref{tab:pmotions}, Figs. \ref{fig:2014}, 
\ref{fig:SED}, \ref{fig:movprop}). The position of the source coincides 
with the FIR 4j core imaged by \citet{Shimajiri2008} and falls close to 
(but slightly offset from) one of the main emission peaks of the dust 
cloud imaged at 3 mm by ALMA (\citealt{Kainulainen2016}; see 
Fig.~\ref{fig:overlays}, right). No counterparts at other wavelengths 
had been reported so far, but the source has been recently imaged at 870 
$\mu$m with ALMA (J. Tobin et al. 2017b, in preparation). The radio 
source has a positive spectral index ($\alpha=1.02\pm0.15$) and we 
suggest that it traces a very young embedded stellar object associated 
with the OMC-2 star-forming region that deserves further study at other 
wavelengths.

\subsection{Proper Motions of the Detected Sources} \label{pmotions}

We searched for proper motions by comparing the positions of the radio 
sources detected in the high-angular resolution maps obtained from 
archive data (epochs 1994-2000) and from our last observations (epoch 
2015). Unfortunately, the sensitivity of the archive data in a single 
epoch was insufficient, and we combined in the same map data taken over 
$\sim$6 years, which is not optimal to measure accurate proper motions, 
because of motions and changes in morphology of the sources over this 
period. Also, the archive data were observed in X-band (8.4-8.5 GHz) 
while the new data have been obtained in C-band (4-8 GHz) (Table 
\ref{tab:observations}). Although the frequency ranges of the two bands 
are almost contiguous, the differences in frequency could 
result in apparent motions due to 
variations of opacity with frequency, particularly in extended sources. 
Despite all these factors that contribute to the uncertainty in our 
measurements, the relatively high velocity of the jet and long time 
($\sim18$ yr) elapsed between the two maps, make the proper motions 
clearly detectable. Our results show that, while VLA 11, and VLA 15 
remain static within the uncertainties, the VLA 12N, VLA 12C and VLA 12S 
knots present measurable proper motions of $\sim$ 40-100 km s$^{-1}$ 
roughly in the S-SW direction (Table \ref{tab:pmotions} and Fig.\ 
\ref{fig:movprop}; see also left panel in Fig.\ \ref{fig:overlays}). 
HOPS 108 shows a measurable proper motion ($32\pm14$ km s$^{-1}$), but 
in the NE direction, whose nature is uncertain (see below). The proper 
motion velocity of VLA 11, the source with the best signal-to-noise 
ratio, is $2\pm2$ km s$^{-1}$ (Table \ref{tab:pmotions}). Since this 
velocity is obtained from the absolute positions of the source, its 
small value indicates that systematic errors are effectively very small 
and that their possible residual effects on the proper motions 
determination are at a level of $\sim$ 2 km s$^{-1}$ or smaller.

\subsubsection{The HOPS 370 Radio Jet} \label{HOPS370}

As discussed above, the knots of VLA 12 are characterized by negative 
spectral indices, indicating non-thermal emission (see Fig. 
\ref{fig:SED}), but their overall morphology with three aligned knots, 
suggests that they pertain to a radio jet. Radio jets associated with YSOs 
are characterized by thermal (free-free) emission (e.g., 
\citealt{Anglada1996}), but in a few cases, non-thermal emission has been 
found in the jet lobes, at relatively large distances from the jet core 
that shows a positive spectral index and thermal emission 
(\citealt{CarGon2013}, \citealt{Anglada2015}, \citealt{RodKam2016} and 
references therein). The most spectacular case of this selected sample of 
non-thermal radio jets is the HH 80 radio jet, where linearly polarized 
emission, indicative of synchrotron radiation, was imaged in the jet lobes 
\citep{CarGon2010}, while the central core of the jet, with a positive 
(thermal) spectral index, remains unpolarized. As the velocities of 
protostellar jets are not relativistic ($<1000$ km~s$^{-1}$), it is 
assumed that the synchrotron emitting electrons are accelerated up to 
relativistic velocities via the diffusive shock acceleration (DSA) 
mechanism \citep{Drury1991} working in the strong shocks present in the 
interaction of the jet with the surrounding medium. Thus, it is expected 
that the VLA 12 knots are associated with strong shocks, as indeed is 
suggested by the detection of several shock tracers in the proximity of 
this source (\citealt{Shimajiri2008}, \citealt{Manoj2013}, 
\citealt{GonGar2016}; see Fig. \ref{fig:overlays}, left panel). Actually, 
the region near HOPS 108 shows the brightest far-IR [OI], CO, H$_2$O 
and OH line emission among the HOPS spectroscopic sample of sources (P. 
Manoj et al. 2017, in preparation), indicating the presence of strong 
shocks.

These results are consistent with the VLA 12 knots being part of a 
non-thermal radio jet either originating from the Class 0 protostar HOPS 
108 (FIR 4), which falls in between the VLA 12C and the VLA 12S knots, 
or from the intermediate-mass Class I YSO HOPS 370 (FIR 3), as the VLA 
12 knots are roughly aligned in the direction of the VLA 11 radio jet 
associated with HOPS 370 (Fig. \ref{fig:2014}) and fall on the SW lobe 
of the submm and far-IR outflow (\citealt{Manoj2013}, 
\citealt{GonGar2016}) driven by this object.

If HOPS 108 was the driving source of the VLA 12 knots, one would expect 
proper motions of these knots away from HOPS 108. In particular, the VLA 
12N and 12C knots would show proper motions with a component pointing to 
the north. However, as Figure \ref{fig:movprop} shows, the proper 
motions of the VLA 12 knots point to the south, consistent with an 
origin from HOPS 370 (VLA 11), and excluding HOPS 108 as the driving 
source. Thus, we conclude that the VLA 12 non-thermal radio knots belong 
to a radio jet, driven by the intermediate-mass Class I object HOPS 370, 
whose central thermal region is traced by the collimated radio source 
VLA 11. VLA 11 is bipolar, extending $\sim1''$ both to the NE and to the 
SW of the position of the protostar (see top-right panel in Fig. 
\ref{fig:movprop}). However, only the SW side of the radio jet shows a 
distant ($\sim30''$) non-thermal lobe (VLA 12), without a NE 
counterpart. The absence of a NE non-thermal radio lobe is probably due 
the lack of a dense clump along the NE jet path where a strong shock 
interaction, similar to that observed in the SW lobe, could take place. 
The jet is also bipolar in the far-IR [OI] lines \citep{GonGar2016}, but 
its SW lobe is more extended, with the brightest part associated with 
the non-thermal VLA 12 radio lobe.

 Adopting 1997.65 as an average epoch for the archive data map, the 
resulting plane-of-the-sky velocities of the VLA 12N, 12C, and 12S knots 
are $\sim$100, 101, and 37 km s$^{-1}$, respectively. The velocities of 
knots 12N and 12C fall in the range of values measured for other 
protostellar radio jets (\citealt{Anglada2015} and references therein); 
the velocity of VLA 12S is smaller, suggesting that this distant knot 
could have suffered a significant deceleration, particularly if the jet 
has interacted strongly with the ambient medium, as indicated by other 
observations (\citealt{Manoj2013}, \citealt{GonGar2016}). In this 
respect, recent proper motion measurements of the radio knots associated 
with the Herbig-Haro objects HH 80, HH 81, and HH 80N, far away from the 
central source, have shown that these knots have significantly slower 
motions than the radio knots located a few arcsec from the central 
source \citep{Masque2015}. This result has been interpreted as 
indicating strong jet interactions with the ambient cloud that 
significantly slow down the jet material.

\subsubsection{The Origin of HOPS 108} \label{HOPS108}

Since the position of the HOPS 108 protostar falls in the path of the 
VLA 11-VLA 12 radio jet, but it can be ruled out as its driving source, 
an appealing alternative possibility is that the jet is related to the 
origin of HOPS 108.
 It has been proposed that shocks associated with the interaction of 
jets with the surrounding medium could compress the gas and induce local 
instabilities that may trigger star-formation 
\citep[e.g.,][]{Yokogawa2003,Graves2010,DuaCab2011}. In particular, this 
scenario has been proposed for HOPS 108, as it is associated with strong 
shocks (\citealt{Manoj2013}, \citealt{GonGar2016}) and probably 
interacting with a nearby dense core \citep{Shimajiri2008}.
 Interestingly, the shape of the jet in the region of the VLA 12 knots 
appears to follow the eastern edge of a clump of enhanced dust emission, 
as imaged by ALMA at 3 mm (\citealt{Kainulainen2016}; see 
Fig.~\ref{fig:overlays}, right panel). Thus, it seems plausible that the 
formation of the HOPS 108 protostar was triggered by the compression of 
the material in this clump after a strong shock interaction with the VLA 
11-VLA 12 jet, driven by HOPS 370.

The evolutionary status of HOPS 370 and HOPS 108 are also fully 
consistent with the suggestion of the latter being triggered by the 
former. HOPS 370, with a bolometric temperature $T_{\rm bol}=72$ K, has 
been classified as an intermediate-mass young Class I object  
(\citealt{Adams2012}, \citealt{Furlan2016}) and we would expect an age 
(time elapsed since the onset of collapse) of a few times $10^5$ years 
for this object. HOPS 108, with a lower bolometric temperature $T_{\rm 
bol}=38$ K, is supposed to be younger, and it has been classified as a 
Class 0 protostar \citep{Adams2012,Furlan2016}. We can then 
hypothesize that some $10^5$ years ago the jet of HOPS 370 
started to impact the molecular clump (see Fig.~\ref{fig:overlays}, 
right panel) from which HOPS 108 later formed. After an interaction of a 
few times $10^4$ years, enough time had passed to allow the collapse of 
the clump and the start of the protostellar stage of HOPS 108. The 
travel time of the gas from HOPS 370 to HOPS 108 is only 500 years (as 
derived from our measured proper motions) and the delay in the influence 
of the first object on the second can be taken to be much shorter than 
the other timescales. We then have a scenario in which the evolutionary 
status of the sources is consistent with one of them triggering the 
formation of the other.

It should be noted that the MIPS 2297 source also falls on the path of 
the HOPS 370 jet, but closer to the origin. Since MIPS 2297 seems to be 
older than HOPS 108 one could think that it is the result of an earlier 
episode of triggered star-formation. If MIPS 2297 is really a Class II 
object, as proposed by \cite{Megeath2012}, it would be even older than 
HOPS 370, making the possibility of triggered formation inviable. 
However, MIPS 2297 is not well studied yet, and if it happens to be 
younger this possibility cannot be fully discarded. On the other hand, 
\cite{Shimajiri2008} proposed that the formation of HOPS 64 ($\sim6''$ 
NW from HOPS 108) could have also been triggered by the outflow from 
HOPS 370. However, in the recent study by \cite{Furlan2016}, both HOPS 
370 and HOPS 64 are classified as Class I sources, suggesting a similar 
evolutionary stage, which would be in conflict with a triggered 
star-formation scenario for HOPS 64. It is also possible that MIPS 2297 
and/or HOPS 64, which are both visible at shorter wavelengths, lie in 
front of the plane where HOPS 370 and HOPS 108 are located; so, they 
could be close to HOPS 370 and its jet just in projection, but in 
reality be farther away and have formed in OMC-2 independently of this 
source. The new source VLA 16 also falls in the path of the
jet, very close to the southernmost detected knot, VLA 12S. We can also not exclude the possibility that star-formation has been triggered, or will eventually be triggered, in other dust cores reported by \cite{Shimajiri2008} which currently do not show evidence of embedded YSOs. Thus, there are at least four
young stars which could be the result of triggering or affected by the HOPS 370 jet, 
but only for HOPS 108 there is further evidence of ongoing 
shock-interaction.

A puzzling issue in the triggered-formation scenario for HOPS 108 is its 
large proper motion velocity ($\sim$ 32 km s$^{-1}$; Table 
\ref{tab:pmotions}). Such a large velocity implies that it should have 
been acquired very recently ($\la$ 200 yr) in order the star to be 
formed in the proximity (within a few arcsec) of its current position, 
in the region where the shock emission has been identified 
\citep{Manoj2013,GonGar2016}. Otherwise, if this velocity was maintained 
during a longer time interval, it would imply that the HOPS 108 
protostar originated at a distant position from its current location, 
making inviable the triggering hypothesis. Relatively low velocities are 
expected if the protostars are mechanically linked to the molecular gas 
reservoir and thus obtain their velocities directly from the gas they 
are forming in, as proposed by \citet{Stutz2016}, who found typical 
radial velocities with respect to the ambient cloud of $\sim$ 0.6 km 
s$^{-1}$ for protostars and $\sim$ 1.8 km s$^{-1}$ for pre-main-sequence 
stars in the integral-shaped filament (ISF) region in Orion. These 
authors propose a slingshot-like ejection mechanism to account for the 
increase in velocity of pre-main-sequence stars relative to protostars. 
However, this mechanism cannot explain velocities above a few km 
s$^{-1}$. 

In contrast, HOPS 108 presents a much larger proper motion velocity, 
similar to the $\sim$ 27 km s$^{-1}$ of the BN object in the nearby 
Orion BN/KL stellar system \citep{Goddi2011,Rodriguez2017}. Another 
runaway star, V 1326 Ori, with a peculiar large proper motion velocity 
of about 28 km s$^{-1}$ has been recently identified in a radio survey 
of proper motions in the core of the Orion Nebula Cluster 
(\citealt{Dzib2017}). For the BN/KL system, in order to explain both the 
uncollimated explosive molecular outflow (\citealt{Zapata2009}) and the 
stellar velocities, violent (proto-)stellar interactions and the 
subsequent dynamical ejection of said (proto-)stars have been invoked 
\citep{Bally2005,Gomez2008,Bally2017}. This acceleration mechanism would 
require the presence of a binary or multiple stellar system near FIR 
4/HOPS 108 to make it compatible with the triggering scenario.
 \citet{Shimajiri2008} imaged 11 potential cores within a region of 
$\sim20''\times20''$ in the proximity of FIR 4 and interpreted them as 
tracing a young stellar cluster whose formation was triggered by the 
impact of the jet with the ambient cloud. If this was the case, 
interactions within this multiple system could have resulted in the 
recent ejection of HOPS 108. We identify the region imaged by 
\citet{Shimajiri2008} with the clump of dust imaged by ALMA at 3 mm 
(\citealt{Kainulainen2016}; Fig.~\ref{fig:overlays}, right panel),  
where the sources HOPS 108, HOPS 64, VLA 15 and VLA 16 appear 
to be embedded. Since HOPS 108 moves toward the NE, 
we do not find a suitable candidate that might be receding from a close 
enough common position and that could be responsible for a past 
interaction with HOPS 108, casting doubt on this possibility.

There are other detected radio sources in the region whose proper 
motions are compatible with being receding from a common location with 
HOPS 108 (e.g., VLA 13, located $\sim40''$ to the SW; A. K. D\'\i 
az-Rodr\'\i guez et al. 2017, in preparation), and that could have 
dynamically interacted with it in the past. However, the location of a 
potential encounter would fall at a large distance from the current 
position of HOPS 108 and, as noted above, would make the origin of this 
object incompatible with a scenario of (local) triggered star-formation. 
On the other hand, if HOPS 108 was accelerated in a distant encounter it 
seems unlikely that our observation occurs just at the moment when it 
crosses the VLA 12 knots of the HOPS 370 jet, at the point of its 
maximum interaction with the ambient cloud.

Another possibility to explain the observed large proper motion of HOPS 
108 is to assume that it is due to a change in the shape of its radio 
emission because of blending with the emission of the nearby VLA 12 
knots. Since these knots move toward the SW, with VLA 12C approaching 
HOPS 108 and VLA 12S going away, a partial blending with these knots can 
result in an apparent motion of HOPS 108 toward the NE.
 Also, a one-sided ejecta of a high velocity cloud of ionized plasma by 
the protostar could produce a change in the shape of its radio emission 
and an apparent proper motion. One-sided high velocity ejecta from young 
stars have been observed in several other sources (e.g., 
\citealt{Rodriguez2017} and references therein). Such an ejecta with a 
velocity of the order of hundreds of km s$^{-1}$ in a poorly resolved 
source would result in a one-sided distortion in the source shape and 
thus a change in the centroid position that could mimic a motion of 
several tens of km s$^{-1}$. Wavelength-dependent opacity effects in an 
extended source could emphasize these changes in the position of the 
centroid of the emission. Indeed, we found that the precise position of 
the HOPS 108 radio source changes with wavelength. Additional, sensitive 
high angular resolution observations could provide a more precise 
measure of the HOPS 108 proper motions and reveal the possible presence 
of either a high velocity ejecta from this source or a nearby 
binary/multiple system whose past interaction with HOPS 108 had been 
responsible for its current high proper motions.

\section{Conclusions}

We analyzed new (2014-2015) multiwavelength (0.7-5 cm), multiconfiguration 
VLA observations together with archive (1994-2000) data at 3.6 cm that 
allowed us to obtain spectral indices and proper motions of radio sources 
in the region enclosing FIR 3 (HOPS 370) and FIR 4 (HOPS 108) in OMC-2. 
Our main conclusions can be summarized as follows:

\begin{enumerate}

\item We detect radio emission from the far-IR sources HOPS 370, HOPS 
108, and HOPS 64, as well as from the mid-IR sources MIPS 2297 and MIPS 
2301. We also detect two new sources, both associated with dust cores, 
that we call VLA 15 and VLA 16. For all these sources we obtain positive 
spectral indices in the centimeter wavelength range, consistent with 
thermal free-free emission from young stellar objects.

\item The radio source VLA 11, associated with HOPS 370, presents a 
clearly elongated bipolar morphology in the NE-SW direction. Because of 
this morphology and positive spectral index ($\alpha \simeq$ 0.3) we 
interpret VLA 11 as the thermal region closest to the origin of a well 
collimated bipolar radio jet driven by the HOPS 370 intermediate-mass 
protostar. We detect several knots of emission (VLA 12N, 12C, and 12S) 
at $\sim$7,500-12,500 au to the SW of HOPS 370 with negative spectral 
indices and showing proper motions of $\sim$40-100 km s$^{-1}$ away from 
the HOPS 370 protostar. We interpret these knots as a non-thermal lobe 
of the HOPS 370 jet. The VLA 12 knots are found in a region where 
previous observations have detected several tracers of strong shocks, 
suggesting that their non-thermal emission is likely synchrotron 
emission from relativistic electrons accelerated in shocks, as has been 
proposed for other non-thermal jet lobes. Although VLA 11 is bipolar, a 
similar non-thermal radio lobe is not found to the NE of HOPS 370, 
probably because there is not a dense clump where the jet interaction 
could take place.

\item HOPS 108 is identified as a compact radio source independent and 
kinematically separated from the radio emission of VLA 12 knots of the 
HOPS 370 jet. Its position along the path of the VLA 12 knots, and 
coincidence with different shock tracers, suggest a scenario where the 
formation of HOPS 108 has been triggered by the interaction of the HOPS 
370 jet with the surrounding medium, as was already proposed by 
\cite{Shimajiri2008}. The more advanced evolutionary stage of HOPS 370 
relative to HOPS 108 and the short dynamical timescale of the jet, are 
consistent with this scenario. However, HOPS 108 presents a large proper 
motion velocity of about 30 km s$^{-1}$ toward the NE whose nature is 
uncertain. This velocity is similar to the proper motion velocities 
found in runaway stellar sources in Orion 
\citep{Rodriguez2017,Dzib2017}. Such a large velocity in HOPS 108 would 
be inconsistent with the triggered scenario, unless the source had been 
accelerated to this high velocity very recently. Alternatively, an 
apparent proper motion could result because of a change in the position 
of the centroid of the source due to a partial blending with the 
emission of the nearby VLA 12 knots or to a one-sided ejecta of ionized 
plasma, rather than by the actual motion of the protostar itself. Deep 
high angular resolution observations at several epochs are required to 
clarify this issue.

\item HOPS 370 and VLA 15 do not show detectable proper motions ($V\la$ 2-4 
km s$^{-1}$), as expected for embedded protostars.

\end{enumerate}

\acknowledgments

A.K.D.R. acknowledges an Spanish MECD FPU fellowship. G.A., A.K.D.R., 
J.F.G., and M.O. acknowledge support from MINECO (Spain) 
AYA2014-57369-C3-3-P grant (co-funded by FEDER). The authors thank 
{\'A}lvaro Sanch\'ez-Monge for helpful suggestions.

\vspace{5mm}
\facilities{VLA}

\software{CASA (v 4.2.2), AIPS}



\vspace{1.5cm}


\floattable
\begin{deluxetable*}{c@{\extracolsep{-0.2em}}c@{\extracolsep{-0.45em}}
c@{\extracolsep{-0.45em}}c@{\extracolsep{-0.18em}}c@{\extracolsep{-0.14em}}
cccc@{\extracolsep{-0.5em}}c@{\extracolsep{-0.3em}}cc@{\extracolsep{-0.26em}}c}[h!]
\tablecaption{Parameters of the VLA Observations 
\label{tab:observations}}
\tabletypesize{\scriptsize}
\tablenum{1}
\tablewidth{0pt}
\tablehead{\colhead{} & \colhead{} & \colhead{} & \colhead{Central} & \colhead{} & \colhead{} & \colhead{} & & &\colhead{Adopted} & \colhead{} & \colhead{Boostrapped} & \colhead{On-Source}  \\
\colhead{} & \colhead{VLA} & \colhead{} & 
\colhead{Frequency} & Bandwidth &
&
\multicolumn{2}{c}{Phase Center} &
\colhead{Flux} & \colhead{Flux Density} & \colhead{Phase} & 
\colhead{Flux 
Density} & \colhead{Time} \\
\cline{7-8}
\colhead{Date} & \colhead{Conf.} & \colhead{Band} & \colhead{(GHz)} & \colhead{(GHz)} & \colhead{Project} & \colhead{$\alpha$(J2000)} &
\colhead{$\delta$(J2000)} & \colhead{Calibrator} & \colhead{(Jy)} & 
\colhead{Calibrator} & \colhead{ (Jy)} & \colhead{(hours)}
}
\startdata
1994 Nov 17 & C & X & 8.4649 & 0.100 & AR0323S &05 35 25.825 & $-$05 09 51.38 & 3C286 & 5.06 &J0541-0541 & 1.240$\pm$0.010 & 0.5\\
1998 Jan 13 & D & X & 8.4851 & 0.100 & AR0387 &05 35 24.220 & $-$05 10 07.27 & 3C48 & 3.15 &J0541-0541 & 0.7420$\pm$0.0010 & 1.0\\
2000 Jan 14 &B & X & 8.4851 & 0.100 & AR0411 &05 35 26.970 & $-$05 10 01.20 & 3C48 & 3.28 &J0541-0541 & 1.299$\pm$0.003 & 3.2\\
2014 Oct 16 & C & K & 22.000 & 8.048 & AM1313 &05 35 27.070 & $-$05 10 00.60 & 3C147 & 2.78 &J0541-0541 & 0.7188$\pm$0.0010& 1.4\\
2014 Oct 16 & C & Q & 44.063 & 8.048 & AM1313 &05 35 27.070 & $-$05 10 00.60 & 3C147 & 0.91 &J0541-0541 & 0.5234$\pm$0.0006& 0.6\\
2014 Oct 18 & C & C & 6.000 & 4.048 & AM1313 &05 35 27.070 & $-$05 10 00.60 & 3C147 & 7.94 &J0541-0541 & 0.9260$\pm$0.0013& 0.5\\
2014 Oct 18 & C & X & 10.000 & 4.048 & AM1313 &05 35 27.070 & $-$05 10 00.60 & 3C147 & 4.84 &J0541-0541 & 0.9050$\pm$0.0006& 0.5\\
2015 Aug 06 & A & C & 6.224 & 3.998 & AO0316 &05 35 27.080 & $-$05 10 00.30 & 3C147 & 7.94 & J0541-0541 & 1.161$\pm$0.004& 0.5\\
2015 Sep 11 & A & C & 6.224 & 3.998 & AO0316 &05 35 27.080 & $-$05 10 00.30 & 3C147 & 7.94 & J0541-0541 & 1.066$\pm$0.004& 0.5\\
\enddata
\end{deluxetable*}


\floattable
\begin{deluxetable*}{ccccll}
\tablecaption{Positions of the Radio Sources \label{tab:positions}}
\tabletypesize{\scriptsize}
\tablenum{2}
\tablewidth{0pt}
\tablehead{ 
\colhead{} & \colhead{Alternative} & \colhead{} & \colhead{} & \multicolumn{2}{c}{Position\tablenotemark{a}} \\
\cline{5-6}
\colhead{Source} & Names & Nature & Ref. & \colhead{$\alpha$(J2000)} & \colhead{$\delta$(J2000)}
}
\startdata
VLA 11 &FIR 3, HOPS 370, MIPS 2302, MIR 21& Class I YSO &1&05 35 27.6337$\pm$0.0007 &$-$05 09 34.368$\pm$0.011 \\
MIPS 2301 & MIR 22 & YSO & 2 &05 35 27.639$\pm$0.008  &$-$05 09 37.09$\pm$0.12 \\
MIPS 2297  & MIR 23 & Class II YSO & 2 &05 35 27.4711$\pm$0.0020 &$-$05 09 44.06$\pm$0.03 \\
VLA 12N &  & Jet knot  & 3 &05 35 27.385$\pm$0.011 &$-$05 09 50.68$\pm$0.16 \\
HOPS 64 & MIPS 2293, MIR 24, FIR 4d/e  & Class I YSO & 1 &05 35 26.9960$\pm$0.0023 &$-$05 09 54.06$\pm$0.03 \\
VLA 12C &  & Jet knot & 3 &05 35 27.111$\pm$0.012  &$-$05 09 58.57$\pm$0.18 \\
HOPS 108 & FIR 4, MIPS 2289 & Class 0 YSO & 1 &05 35 27.086$\pm$0.004  &$-$05 09 59.95$\pm$0.06 \\
VLA 12S &  & Jet knot & 3 &05 35 26.9389$\pm$0.0017 &$-$05 10 02.876$\pm$0.025 \\
VLA 16 & & YSO & 3 &05 35 26.824$\pm$0.003 &$-$05 10 05.64$\pm$0.04 \\
VLA 15  & FIR 4j & YSO & 3 &05 35 26.4091$\pm$0.0011 &$-$05 10 05.951$\pm$0.016 \\
\enddata
 \tablenotetext{a}{Positions derived from elliptical Gaussian fits in 
the A-configuration C-band map of epoch 2015, except for MIPS 2301 and 
MIPS 2297 that are not detected in this map, and whose positions were 
derived from the 1994-2000 X-band and the 2014 Q-band maps, 
respectively. 
 Uncertainties correspond to those of absolute positions, and are 
calculated adding in quadrature a systematic error of 0\farcs01 
\citep{Dzib2017} to the formal error of the fit, $0.5\,\theta / {\rm 
SNR}$ \citep{Reid1988}. The systematic error accounts for uncertainties 
introduced by the phase calibration process and the error of the fit for 
those due to the source size and signal-to-noise ratio.
 }
 \tablerefs{(1) \citealt{Furlan2016}; (2) \citealt{Megeath2012}; (3) 
This paper.}
  \end{deluxetable*}
 
   
\floattable
\begin{deluxetable*}{ccccccccccc}
\tablecaption{Flux Densities and Proper Motions of the Sources \label{tab:pmotions}}
\tabletypesize{\scriptsize}
\tablenum{3}
\tablewidth{0pt}
\tablehead{
\colhead{} & \multicolumn{5}{c}{Flux Density\tablenotemark{a}} & \colhead{} & \multicolumn{4}{c}{Proper Motions\tablenotemark{d}} \\ 
\cline{2-6} \cline{8-11} 
\colhead{} & \colhead{5 cm} & \colhead{3 cm} & \colhead{1.3 cm} & \colhead{0.9 cm\tablenotemark{b}} & \colhead{0.7 cm} & 
\colhead{Spectral} &
\colhead{$\mu_{\alpha} \cos \delta$} &
\colhead{$\mu_{\delta}$} & 
\colhead{$V$} &
\colhead{P.A.} \\
\colhead{Source} &\colhead{(mJy)} &\colhead{(mJy)} & \colhead{(mJy)} & \colhead{(mJy)} &
\colhead{(mJy)} & \colhead{Index\tablenotemark{c}} &\colhead{(mas yr$^{-1}$)} & \colhead{(mas
yr$^{-1}$)} &\colhead{(km s$^{-1}$)} &\colhead{(deg)} 
}
\startdata
VLA 11  &1.76$\pm$0.18 &2.16$\pm$0.22 &2.6$\pm$0.3 &2.6$\pm$0.3 &4.0$\pm$0.4 &0.27$\pm$0.10 &0.7$\pm$0.8  &0.7$\pm$0.8  &2$\pm$2  &50$\pm$50\\
MIPS 2301 & 0.016$\pm$0.004\tablenotemark{e} &0.032$\pm$0.010\tablenotemark{f} &$<$ 0.03 &$<$ 0.03 &$<$ 0.07 &0.30$\pm$0.23 &\nodata\tablenotemark{g} 
&\nodata\tablenotemark{g} &\nodata\tablenotemark{g} &\nodata\tablenotemark{g} \\
MIPS 2297 &$<$ 0.03 &$<$ 0.03 &0.049$\pm$0.012  &0.092$\pm$0.013  &0.116$\pm$0.022  &1.4$\pm$0.5 &\nodata\tablenotemark{g}  &\nodata\tablenotemark{g}  &\nodata\tablenotemark{g}  
&\nodata\tablenotemark{g}\\
VLA 12N &0.134$\pm$0.020  &0.073$\pm$0.014  &$<$ 0.03 &$<$ 0.025 &$<$ 0.05 &$-$1.07$\pm$0.07  &22$\pm$11 &$-$46$\pm$11  &100$\pm$21 &155$\pm$12\\
HOPS 64 &0.011$\pm$0.003\tablenotemark{e} &$<$ 0.03 &0.069$\pm$0.012 &0.140$\pm$0.016  &0.30$\pm$0.03 &1.51$\pm$0.06 &\nodata\tablenotemark{g}  &\nodata\tablenotemark{g}  
&\nodata\tablenotemark{g}  &\nodata\tablenotemark{g}\\
VLA 12C &0.24$\pm$0.03\tablenotemark{h} &0.09$\pm$0.03\tablenotemark{h} &$<$ 0.03 &$<$ 0.024 &$<$ 0.04  &$-$1.3$\pm$0.4 &$-$11$\pm$14  &$-$50$\pm$15  &100$\pm$30 &192$\pm$16\\
HOPS 108  &0.018$\pm$0.003\tablenotemark{e} &0.060$\pm$0.014\tablenotemark{f} &0.047$\pm$0.011  &0.115$\pm$0.014  &0.084$\pm$0.017 &0.7$\pm$0.3 &14$\pm$7 &9$\pm$7 &32$\pm$14 
&57$\pm$25\\
VLA 12S &0.25$\pm$0.03 &0.192$\pm$0.024  &0.16$\pm$0.03  &0.137$\pm$0.020  & $<$ 0.04 &$-$0.59$\pm$0.20 &$-$13$\pm$3  &$-$14$\pm$3  &37$\pm$6  &223$\pm$10\\
VLA 16 &0.017$\pm$0.003\tablenotemark{e} &$<$ 0.03 &0.040$\pm$0.010 &0.059$\pm$0.012 &0.069$\pm$0.014 &0.73$\pm$0.06 
&\nodata\tablenotemark{g} &\nodata\tablenotemark{g} 
&\nodata\tablenotemark{g} &\nodata\tablenotemark{g}\\
VLA 15 &0.071$\pm$0.014  &0.100$\pm$0.013  &0.200$\pm$0.022  &0.32$\pm$0.04 &0.60$\pm$0.07 &1.02$\pm$0.15 &$-$1$\pm$3 &2$\pm$3 &4$\pm$6 &340$\pm$90 \\ 
\enddata
 \tablenotetext{a}{Measured in the 2014 C-configuration maps, except when noted. Primary beam correction has been applied. Upper limits are 3\,$\sigma$.} 
 \tablenotetext{b}{From C-configuration observations made on 27 March 
2016 (J. Tobin et al. 2017a, in preparation).}
 \tablenotetext{c}{The datapoint at 0.7 cm has been excluded from the 
fit in the sources with positive spectral index to avoid possible 
contamination from dust emission.} 
 \tablenotetext{d}{Measured from the X-band map made from 1994-2000 data 
and the C-band map at epoch 2015. Positions were obtained from 
elliptical Gaussian fits. Positional uncertainties were calculated 
as explained in the footnote of Table \ref{tab:positions}.
 Errors in the proper motions were estimated from the uncertainties in 
the absolute positions using standard propagation error theory.
 For resolved sources, opacity effects can produce small differences in
the positions if obtained from maps at different wavelengths and/or
angular resolutions, resulting in an additional uncertainty in the
proper motion measurements.}
  \tablenotetext{e}{Obtained from the C-band 2015 A-configuration map 
because the source is not detected or appears blended with nearby 
sources in the C-band 2014 C-configuration map.}
 \tablenotetext{f}{Obtained from the 1994-2000 multiconfiguration map at 
3.6 cm because the source appears blended with nearby sources in the 
X-band 2014 C-configuration map.}
 \tablenotetext{g}{Not detected in the first epoch map (1994-2000 data) 
or too weak for a reliable positional fit.}
  \tablenotetext{h}{The expected contribution of HOPS 108, estimated from 
higher resolution data, has been subtracted.}
 \end{deluxetable*}

\newpage
 
\newpage

\newpage

\newpage

\end{document}